\documentclass[a4paper,11pt]{article}
\pdfoutput=1 % if your are submitting a pdflatex (i.e. if you have
\usepackage{jheppub} % for details on the use of the package, please
\usepackage[T1]{fontenc} % if needed
\usepackage{subcaption}

\usepackage[mathscr]{euscript}
\usepackage{empheq}

\newcommand{\taup}{\mathscr{T}}
\newcommand{\vcut}{v}%v_{\rm cut}

\newcommand{\hyp}{{}_2F_1}

\newcommand{\CF}{C_F}
\newcommand{\CA}{C_A}
\newcommand{\tCF}{\widetilde{C}_F}
\newcommand{\tCA}{\widetilde{C}_A}
\newcommand{\CR}{C_R}
\newcommand{\as}{\alpha_s}

\newcommand{\Ralt}{\mathcal{R}}
\newcommand{\logt}{l}
\newcommand{\logtsq}{L}
\newcommand{\tC}{\widetilde{C}}
\newcommand{\li}[1]{\text{Li}_{#1}}
\newcommand{\CtildeCH}{\tC^{({\chi^2})}}
\newcommand{\CtildeXE}{\tC^{({\rm XE})}}
\newcommand{\intCH}{\mathcal{I}^{({\chi^2})}}
\newcommand{\intXE}{\mathcal{I}^{({\rm XE})}}
\newcommand{\BAC}{\mathcal{B}}
\newcommand{\SIG}{\mathcal{S}}
\usepackage{xspace}
\newcommand{\nsub}{$N$-subjettiness\xspace}

\usepackage{multirow}
\usepackage{pifont}

\usepackage{xcolor}
\definecolor{darkgreen}{rgb}{0.0, 0.5, 0.0}

\title{\boldmath  Towards Machine Learning Analytics \\ for Jet Substructure}
\author[a]{Gregor Kasieczka,}
\author[b]{Simone~Marzani,} 
\author[c]{Gregory~Soyez,}
\author[d,e,f]{Giovanni~Stagnitto.}

\affiliation[a]{Institut f\"ur Experimentalphysik, Universit\"at Hamburg,\\Luruper Chaussee 149, D-22761 Hamburg, Germany}
\affiliation[b]{Dipartimento di Fisica, Universit\`a di Genova and INFN, Sezione di Genova,\\ Via Dodecaneso 33, 16146, Genoa, Italy}
\affiliation[c]{Institut de Physique Th\'eorique, Paris Saclay University, CEA, CNRS, F-91191 Gif-sur-Yvette France}
\affiliation[d]{Sorbonne Universit\'e, CNRS, Laboratoire de Physique 
Th\'eorique et Hautes \'Energies,\\ LPTHE, F-75005 Paris, France}
\affiliation[e]{Universit\'e de Paris, LPTHE, F-75005 Paris, France}
\affiliation[f]{Tif Lab, Dipartimento di Fisica, Universit\`a di Milano 
and INFN, Sezione di Milano,\\ Via Celoria 16, I-20133 Milano, Italy}

\emailAdd{gregor.kasieczka@uni-hamburg.de}
\emailAdd{simone.marzani@ge.infn.it}
\emailAdd{gregory.soyez@ipht.fr}
\emailAdd{giovanni.stagnitto@lpthe.jussieu.fr}

\preprint{
\begin{flushright}
TIF-UNIMI-2020-12
\end{flushright}
}

\abstract{ 
The past few years have seen a rapid development of machine-learning algorithms.
While surely augmenting performance, these complex tools are often treated as black-boxes and may impair our understanding of the physical processes under study.
The aim of this paper is to move a first step into the direction of applying expert-knowledge in particle physics to calculate the optimal decision function and test whether it is achieved by standard training, thus making the aforementioned black-box more transparent.  
In particular, we consider the binary classification problem of discriminating quark-initiated jets from gluon-initiated ones. We construct a new version of the widely used \nsub, which features a simpler theoretical behaviour than the original one, while maintaining, if not exceeding, the discrimination power. We input these new observables to the simplest possible neural network, i.e.\ the one made by a single neuron, or perceptron, and we analytically study the network behaviour at leading logarithmic accuracy. We are able to determine under which circumstances the perceptron achieves optimal performance. We also compare our analytic findings to an actual implementation of a perceptron and to a more realistic neural network and find very good agreement.

}

\begin{document} 
\maketitle
\flushbottom

\section{Introduction}\label{sec:intro}

The CERN Large Hadron Collider (LHC) is continuously exploring the high-energy frontier. However, although well-motivated from a theoretical viewpoint, no conclusive evidence of new particles or interactions, and no significant cracks in the Standard Model (SM) of particle
physics have been found. This challenges the particle physics community, theorists and experimenters alike, to find new ways to interrogate and interpret the data.
In the context of analyses involving hadronic final states, jet substructure has emerged as an important tool for searches at the LHC. Consequently, a vibrant field of theoretical and experimental research has developed in the past ten years, producing a variety of studies and techniques~\cite{Marzani:2019hun}.
Many grooming and tagging algorithms have been developed, successfully tested, and are currently used in experimental analyses~\cite{Abdesselam:2010pt,Altheimer:2012mn,Altheimer:2013yza,Adams:2015hiv,Larkoski:2017jix,Asquith:2018igt}.

The field of jet substructure is currently undergoing a revolution. The rapid development, within and outside academia, of machine-learning (ML) techniques is having a profound impact on particle physics in general and jet physics in particular. 
In the context of high-energy physics, ML classification algorithms are typically
trained on a control sample, which could be either Monte Carlo
pseudo-data or a high-purity dataset, and then applied to an unknown sample to classify its properties. 
These ideas have been exploited in particle physics for a long time. However, because of limitations on the algorithms' efficiency and on computers' power, so-far algorithms were applied to relatively low-dimensional projections of the full radiation pattern that one wished to classify. Even so, such projections usually correspond to physically-motivated observables, such as, e.g.\ the jet mass, jet shapes, and therefore limitation in performance was mitigated with physics understanding. 
The current ML revolution moves away from low- dimensional projections and exploit deep neural network (NN) to perform classification. 
Early progress was made on supervised classification of 
known particles~\cite{jet_images2,jets_w,jets_qg2,jets_top,deep_top1}
to the point where powerful network architectures are now 
available~\cite{jets_comparison,particlenet,jedinet,DeepJet,Komiske:2018cqr}
and the focus lies on improving the stability~\cite{pivot,disco,bayes,bayes2,ben_unc}
of these approaches.

These techniques, developed by both theorists and experimentalists, have already been tested in the LHC environment and dedicated studies have been performed by the ATLAS and CMS collaborations~\cite{ATLASTop,ATLASCorrelation,CMSTop}. Despite its rapid development and the unquestionable improvements that ML brings to particle physics, it is often met with a certain degree of suspicion, because physicists often feel uneasy about relying on black-box techniques. 
We believe that the situation is not too dissimilar to the one we
faced a decade ago, at the birth of jet substructure. At the time
there was a plethora of new tagging algorithms and a deeper
understanding was reached only when theoretical studies based on QCD
were performed, see e.g.~\cite{Dasgupta:2013ihk,Dasgupta:2013via,Larkoski:2014wba, Dasgupta:2015lxh,Salam:2016yht, Dasgupta:2016ktv,Dasgupta:2015yua,Larkoski:2014gra,Larkoski:2015kga,Larkoski:2020wgx, Larkoski:2013eya,Kang:2019prh,Cal:2019gxa}. 

A common expectation is that well-trained networks can achieve the performance of the likelihood ratio, i.e.\ they can learn a monotonic function of the likelihood ratio.
In this paper, we are in the almost-unique
position to analytically derive the weights corresponding to  
a cut on the likelihood ratio for a realistic problem and to test whether they agree with a network trained the
usual way.
Inspired by the successful approach of the past, we want to anchor
this investigation in perturbative QCD, with a first-principle
understanding approach, by looking at a situation where most studies can be
performed analytically, and afterwards by investing how our findings in the
simplest case compare to a more general setup.

For this current study, we focus on an issue that has received a great
deal of both theoretical and experimental attention, namely the
ability of disentangling jets that can be thought as having originated
by the fragmentation of a high-energy quark (henceforth quark jets),
from the ones originated from a gluon (henceforth gluon jets)~\cite{Gras:2017jty}.
In particular, in section~\ref{sec:newtau}, we introduce a variant of the well-known \nsub
variables~\cite{Thaler:2010tr} based on the primary
Lund plane declustering tree~\cite{Dreyer:2018nbf}.
This {\em primary} \nsub $\taup_N$ is more amenable to an all-order
QCD analysis, which we perform at leading logarithmic (LL) accuracy,
while maintaining, if not improving, the quark/gluon discrimination power. 
This definition is such that, if we measure the \nsub variables $\{\taup_1...\taup_n\}$, then, at LL, a cut on the likelihood ratio simply corresponds to a cut on $\taup_n$. 
With this definition, and within the stated accuracy, we are able to
write simple analytic expressions for the cumulative distributions and for the
so-called area under the curve (AUC), which measures the
discriminating power of the observable,  for any value of $n$.~\footnote{Expressions for cumulative distributions
  can also be obtained for the standard \nsub definition, see e.g.~\cite{Dasgupta:2015lxh,Larkoski:2019nwj}, although their
  complexity increases with $n$ to a point where other quantities, such as, for instance,
  the AUC are no longer tractable analytically.}

We then move in section~\ref{sec:analytics} to use the aforementioned
primary \nsub variables as input to a NN. Because we expect the
optimal discriminant to be a cut on $\taup_n$, we can start by
considering the simplest possible NN, namely a single neuron, or {\em
  perceptron}, with $n$ inputs and one output. This opens up the
interesting possibility of performing analytic calculations that
describe the behaviour of the perceptron at LL accuracy in QCD and allow us to determine the optimal decision function.
The ability to fully control the behaviour of
  a ML technique analytically from first-principles is the main novelty
  of this paper.
In sections~\ref{sec:numerics} and~\ref{sec:montecarlo}, we compare
our theoretical findings with actual networks with architecture of
increasing complexity using pseudo-data,  which we generate either
according to a leading-logarithmic distribution in QCD
(section~\ref{sec:analytics}) or using a general-purpose Monte Carlo
parton shower (section~\ref{sec:montecarlo}). Finally, we draw our conclusions in section~\ref{sec:conclusions}.

\section{Quark/gluon tagging with \nsub}\label{sec:newtau}

In the framework of perturbative QCD, final-state jets, i.e.\ collimated sprays of hadrons, are formed by the fragmentation of high-energetic partons, which emerge from the hard scattering. While the association of a single parton, say a quark or a gluon, to a final-state jet is an intrinsically ambiguous --- if not ill-defined --- operation, essentially because of higher-order corrections --- it is however possible to employ operational definitions that associate the value of a measurable quantity to an enriched sample of quarks or gluons~\cite{Badger:2016bpw,Gras:2017jty}.
Many such observables have been proposed in the literature, ranging from more traditional jet observables such as angularities~\cite{Larkoski:2014pca} or energy-correlation functions~\cite{Larkoski:2013eya}, to observables that exploit more modern jet substructure techniques, such us the iterated soft-drop multiplicity~\cite{Frye:2017yrw} or the Les Houches multiplicity~\cite{Amoroso:2020lgh}, to operational definitions inspired by data science~\cite{Metodiev:2018ftz,Komiske:2018vkc}.

A general approach to quantitatively assess the power of quark/gluon discrimination was put forward in Ref.~\cite{Larkoski:2019nwj}. The core idea of that approach is to measure multiple infra-red and collinear (IRC) safe observables on a single jet as a means to characterise the available emission phase space. 
The observables of choice were the \nsub variables~\cite{Thaler:2010tr,Thaler:2011gf} $\tau_N$, where the measurement of the first $n$ variables $\{\tau_1 \dots \tau_n\}$ is able to resolve $n$ emissions in the jet\footnote{There is indeed more information in a jet than what is captured by this set of \nsub variables. One could study, for instance, an extended set of \nsub variables~\cite{Datta:2017rhs}, energy-flow polynomial~\cite{Komiske:2017aww}, or counting observables, e.g.~\cite{Frye:2017yrw}. We thank Andrew Larkoski and Ben Nachman for comments on this point.}.
% variables} 
%
The \nsub shape is one of the most used jet substructure variables. As its name suggests, it aims to identify jets with an $N$-prong structure, taking inspiration from the event-shape $N$-jettiness~\cite{Stewart:2010tn}.
In order to achieve this, a set of axes $a_1,\dots,a_N$ is introduced and the \nsub is defined as   
\begin{equation}\label{eq:Nsubjettiness}
\tau_N^{(\beta)} = \frac{1}{p_{t} R_0^\beta}\sum_{i\in\text{jet}} p_{ti}\,
  {\text{min}}(\Delta_{ia_1}^\beta,\dots,\Delta_{ia_N}^\beta),
\end{equation}
where $\beta$ is a free parameter often set to unity, $p_{t}$ and $R_0$ are the jet transverse momentum and radius respectively and $\Delta_{ia_j}=\sqrt{\Delta y_{ia_j}^2+\Delta\phi_{ia_j}^2}$ is the distance between particle $i$ and the axis $a_j$ in the azimuth-rapidity plane.
Note that the axes $a_j$  can be defined in several ways, for a review see for instance Ref.~\cite{Marzani:2019hun}.
Crucially, IRC safety guarantees that \nsub distributions can be meaningfully calculated in perturbative QCD. 
These distributions have been the subject of several theoretical investigations~\cite{Dasgupta:2015lxh,Salam:2016yht,Napoletano:2018ohv} (see also~\cite{Larkoski:2015kga,Larkoski:2017cqq,Moult:2017okx} for theoretical studies on the related observable $D_2$).
The \nsub distributions were calculated explicitly at LL accuracy, i.e.\ in the limit in which all emissions are strongly ordered, in Ref.~\cite{Larkoski:2019nwj} for the cases $n=1,2,3$. Furthermore, the Authors of Ref.~\cite{Larkoski:2019nwj} were able to calculate from first-principle familiar quantities that enter statistical analyses. 
We start our discussion with a brief recap of the elements in Ref.~\cite{Larkoski:2019nwj} which are relevant for our current study.

The likelihood ratio, which according to the  Neyman-Pearson lemma~\cite{Neyman:1933wgr} provides the best single-variable discriminant, is simply given by the ratio of the probability distributions computed for background (gluons) and signal (quark) jets: 
\begin{equation}\label{likelihood-def}
\mathcal{L}(\tau_1,\dots, \tau_n)= \frac{p_\mathcal{B}}{p_\mathcal{S}}=\frac{p_g(\tau_1,\dots, \tau_n)}{p_q(\tau_1,\dots , \tau_n)},
\end{equation}
where
\begin{equation}\label{distr-def}
p_{i}(\tau_1, \dots, \tau_n) =
\frac{1}{\sigma_{i}}\frac{d \sigma_{i}}{d\tau_1 \cdots d\tau_n}, \quad i=q,g.
\end{equation}
%\item 
In order to assess the discriminating power of an observable, Receiver
Operating Characteristic (ROC) curves are often considered. These
curves show the background (gluon) efficiency against the signal
(quark) efficiency and are remarkably useful  to directly compare the
performance of different tools. Furthermore, note that ROC curves are invariant under monotonic rescaling of the inputs.

For a single-variable observables $V\equiv V(\tau_1,\dots,\tau_n)$,
one can define the normalised cumulative distribution
\begin{equation}\label{eq:sigma-def}
  \Sigma_i(v)=\int d\tau_1\dots d\tau_n\,
  p_i(\tau_1,\dots,\tau_n)\,\Theta(V(\tau_2,\dots,\tau_n)<v).
\end{equation}
The ROC curve is then obtained as
\begin{equation}\label{roc-def}
\text{ROC}(x) = \Sigma_g\left (\Sigma_q^{-1}\left(x \right)\right)
\end{equation}
where $x$ is the signal (quark) efficiency.
The area under the ROC curve (AUC) can be used as a quantifier of the
discrimination power, where $\text{AUC}=0$ corresponds to perfect performance.\footnote{Note than one could have
  alternatively defined the ROC curve as showing the background
  rejection against the signal efficiency i.e.\
  $\text{ROC}(x) = 1 - \Sigma_g\left (\Sigma_q^{-1}\left(x
    \right)\right)$. In this case the perfect performance would
  correspond to $\text{AUC}=1$.}
For example, if one considers an IRC observable $v$ that only resolves one emission, e.g.\ $v=\tau_1$, then the LL cumulative distributions for quarks and gluons simply differ by their colour factor, leading to the so-called Casimir scaling:
\begin{align}
\text{ROC}(x) =  \left(x\right) ^\frac{\CA}{\CF}, \quad \text{AUC}=\int_0^1 dx\, x^\frac{\CA}{\CF}=\frac{\CF}{\CF+\CA}.
\end{align}

The fact that $\tau_n$ resolves $n$ emissions inside the jet results
in a rather complicated structure, even in the limit where the
emissions are strongly ordered. This is because the emissions that set
the values of the observables $\tau_i$,  which are always gluons at LL
accuracy, can either be primary emissions, i.e.\ they originate from
the original hard parton which could be a quark or a gluon, or they
can originate from subsequent gluon splittings. If we consider, for
instance, the case of a jet initiated by a hard quark, one ends up
with $n$ contributions with colour factors $\CF^{n-i} \CA^i$,
$i=0,\cdots ,n-1$.
Furthermore, a rather intricate resummation structure emerges, 
i.e.\ a Sudakov form factor with both $C_F$ and $C_A$ contributions
and depending on the complete tower of $n$ emissions.
It is clear that this intricate structure does not facilitate
analytical calculations, especially because, in this study, we are not
interested in considering $p_q$ and $p_g$ as final results, but
rather, as inputs for subsequent calculations.

As a consequence, we find convenient to introduce a variant of \nsub that is sensitive, at LL accuracy, only to primary emissions, such that the distributions $p_i$ are determined by strongly-ordered gluon emissions off the initial hard parton. We present this new observable in the next section.

\subsection{Primary \nsub} 

We define the new observable primary \nsub as follows. Starting
from a jet of radius $R_0$, one first
builds the list of primary Lund declusterings~\cite{Dreyer:2018nbf}:
\begin{enumerate}
\item Recluster the jet constituents with the Cambridge/Aachen
  algorithm~\cite{Dokshitzer:1997in,Wobisch:1998wt}. 
\item Iteratively undo the last step of the clustering $j\to
  j_1+j_2$, with $p_{t1}>p_{t2}$. At step $i$ ($i=1,\dots,m$), define
  \begin{equation}
    \tilde{p}_{ti} = p_{t2}
    \qquad\text{ and }\qquad
    \Delta_i = \sqrt{\Delta y_{12}^2 + \Delta \phi_{12}^2}.
  \end{equation}
  Repeat the procedure with $j=j_1$, i.e.\ following the
  harder branch of the declustering.
\item When the declustering terminates, i.e.\ when $j$ is no longer
  coming from a $j_1+j_2$ clustering, define $\tilde p_{t0}$ as the
  transverse momentum of $j$.
\end{enumerate}
From the set of transverse momenta, we can define the momentum fractions
\begin{equation}
  z_i = \frac{\tilde{p}_{ti}}{\sum_{i=0}^m \tilde{p}_{ti}} 
  \qquad i=0,\dots,m,
\end{equation}
where we note that the final hard momentum $\tilde{p}_{t0}$ is
included in the normalisation.
This produces a set of values $(z_i, \Delta_i)$, for $i=1,\dots, m$, that we order such
that $z_1\Delta_1^\beta\ge z_2 \Delta_2^\beta\ge\dots\ge
z_m\Delta_m^\beta$.
The primary \nsub  is then defined as
\footnote{Note that an alternative definition, equivalent at leading-logarithmic
accuracy, but different beyond, would be to define $\taup^\text{(max)}_N=z_N\left(\tfrac{\Delta_N}{R_0}\right)^\beta$.}
\begin{equation}\label{eq:tau-prim-def}
  \taup_N = \sum_{i=N}^m z_i \left(\frac{\Delta_i}{R_0}\right)^\beta.
\end{equation}
Note that $\taup_1 \ge \dots \ge \taup_n$, like with the standard \nsub
$\tau_N$.
The primary \nsub definition in Eq.~(\ref{eq:tau-prim-def}) is very
similar to the standard \nsub  definition with the main difference
that it is computed based on primary Lund declusterings. The
definition of the momentum fractions $z_i$ is such that
$\sum_{i=0}^mz_i=1$.

\subsection{Primary \nsub at leading-logarithmic accuracy}\label{sec:prim-ll-study}

By construction, the leading-logarithmic (LL) expression for the
$n$-dimensional differential distribution is obtained by considering
strongly-ordered independent emissions off the originating hard parton
(either a quark or a gluon).
In this strongly-ordered limit, only the first term in the rhs of
Eq.~\eqref{eq:tau-prim-def} should be kept, i.e.\ $\taup_N\approx
z_N(\Delta_N/R_0)^\beta$.
In that context, the structure of the leading-logarithmic resummation
is particularly trivial: one gets a factor for each of the $n$
emissions associated with $\taup_1,\dots,\taup_n$ as well as a Sudakov
factor vetoing any additional real emission with
$z(\Delta/R_0)^\beta>\taup_n$.\footnote{As usual, this is equivalent
  to saying that real and virtual emissions cancel each other for
  $z(\Delta/R_0)^\beta<\taup_n$ and only virtual corrections
  contribute for $z(\Delta/R_0)^\beta>\taup_n$. These virtual
  corrections trivially exponentiate.}
This yields
\begin{align}\label{eq:primdiff-end}
  p_i(\taup_1, \dots, \taup_n) &=
  \left( \prod_{j=1}^n \frac{ \Ralt'(\taup_j)}{\taup_j}\right) \left( C_i\right)^n \exp\left[-C_i \Ralt(\taup_n) \right],
\end{align}
where $i=q, g$ refers to the flavour
of the jet (with $C_q=\CF$ and $C_g=\CA$).
The radiator $\Ralt$, which and has been stripped
of its colour factor, is
computed in the soft-and-collinear-limit including running-coupling
corrections, as appropriate for our LL accuracy:
\begin{align} \label{eq:radiator}
  \Ralt(\taup) &= 2 \int_0^1 \frac{d\theta^2}{\theta^2} \int_0^1\frac{dz}{z}
  \frac{\alpha_s(z\theta p_t R_0)}{2\pi} \Theta(z\theta^\beta > \taup) 
  = \frac{2}{\beta}\int_\taup^1 \frac{d\theta^\beta}{\theta^\beta} \int_{\taup/\theta^\beta}^1\frac{dz}{z}
                 \frac{\alpha_s(z\theta p_t R_0)}{\pi}. 
\end{align}
We have also introduced 
$\Ralt'(\taup) = \frac{d \Ralt(\taup)}{d \log(1/\taup)}$, where $\log x$ always denotes the natural logarithm of $x$.

An important observation is the following. From Eq.~(\ref{eq:primdiff-end}) we note that the structure of the probability
distributions at LL in QCD for primary definition of \nsub is the {\em same} for quark
and gluon jets except for the colour factor, $C_F$ or $C_A$ which
appears as an overall factor in both the $\Ralt'$ pre-factors and the Sudakov exponent.
(This is not case with the standard definition of \nsub).
Consequently, the likelihood ratio Eq.~(\ref{likelihood-def}) at LL
becomes
\begin{equation}\label{likelihood-LL}
  \mathcal{L}^\text{LL}
   =\left( \frac{\CA}{\CF}\right)^n\exp\left[-(\CA-\CF)
     \Ralt(\taup_n) \right],
 \end{equation}
which is a monotonic function of $\Ralt(\taup_n)$, and hence at LL of $\taup_n$, only. Therefore, at LL, a
cut on the likelihood ratio is equivalent to a cut on $\taup_n$. The
remarkable simplicity of this result is the strongest motivation for
introducing primary \nsub. This observable is thus the ideal
laboratory to study analytically how a neural network that takes the
primary \nsub variables as inputs performs. Due to  the simplicity of
the classifier --- a cut on a single variable --- we expect that even
the simplest network, i.e.\ a single neuron, should be enough to achieve optimal classification performance, i.e.\ a perceptron should be able to
learn a function which is monotonically related to likelihood ratio. 
Studying more complex architectures --- especially including one or more hidden layers --- would be very interesting but is not in the scope of this exploratory work. 

Analytic studies of a perceptron will be the topic of section~\ref{sec:analytics}, but, before moving to that, let us derive a couple of results that allow us to establish primary \nsub as an appealing observable on its own, rather than just a shortcut to more tractable analytic results. 
Firstly, it is interesting to obtain an analytic expression for the cumulative distribution with a cut $\taup$ on $\taup_n$. As we have just seen, this is equivalent to a cut on the likelihood. We have 
\begin{align}\label{cumulative-start}
  \Sigma_i(\taup_n < \taup)
 & = \int_0^1 d\taup_1 \int_0^{\taup_1} d\taup_2 \dots  \int_0^{\taup_{n-2}} d\taup_{n-1} \int_0^{\taup_{n-1}} d\taup_{n}\,p_i(\taup_1, \dots, \taup_n)\,\Theta(\taup_n<\taup) \nonumber \\
  & = \int_0^1 d\taup_1 \int_0^{\taup_1} d\taup_2 \dots  \int_0^{\taup_{n-2}} d\taup_{n-1} \int_0^{\min[\taup_{n-1},\taup]} d\taup_{n}\,p_i(\taup_1, \dots, \taup_n)\,,
\end{align}
where $i=q,g$.
The latter expression splits naturally in two terms: if $\taup_{n-1} <
\taup$, we simply find $\Sigma_i(\taup_{n-1} < \taup)$; if
$\taup_{n-1} > \taup$, the exponential in~(\ref{eq:primdiff-end}) factors out and we obtain
\begin{equation}
  e^{-C_i \Ralt(\taup)} \int_\taup^1 d\taup_1 \int_\taup^{\taup_1} d\taup_2 \dots  \int_\taup^{\taup_{n-2}} d\taup_{n-1}\,p_i(\taup_1, \dots, \taup_n)
  = e^{-C_i \Ralt(\taup)} \frac{C_i^{n-1} \Ralt^{n-1}(\taup)}{(n-1)!}\,.
\end{equation}
By induction we arrive at
\begin{equation}\label{cumulative-res}
  \Sigma_i(\taup_n< \taup) = e^{-C_i \Ralt(\taup)} \sum_{k=1}^n \frac{C_i^{k-1} \Ralt^{k-1}(\taup)}{(k-1)!}
  = \frac{\Gamma(n,C_i \Ralt(\taup))}{\Gamma(n)},
\end{equation}
where $\Gamma(n,x)$ is the incomplete Gamma function.

It is also possible to find an analytic expression for the AUC. Exploiting the  ROC curve definition in Eq.~(\ref{roc-def}), we can write the AUC as an integral of the quark and gluon distributions:
\begin{align} \label{auc-eval}
  \text{AUC} &= \int_0^1 d\taup_{1q} \int_0^{\taup_{1q}} d\taup_{2q} \cdots \int_0^{\taup_{n-1,q}} d\taup_{nq}
  \int_0^1 d\taup_{1g} \int_0^{\taup_{1g}} d\taup_{2g} \cdots \int_0^{\taup_{n-1,g}} d\taup_{ng} \nonumber \\
  & \phantom{\int_0^1} \cdot p_q(\taup_{1q}, \taup_{2q}, \cdots, \taup_{nq})\,p_g(\taup_{1g}, \taup_{2g}, \cdots, \taup_{ng})
  \,\Theta(\taup_{nq} > \taup_{ng})\,.
\end{align}
The details of the integration are not all trivial and are thus given in
appendix~\ref{app:details}. Here we just quote the final result:
\begin{equation}\label{auc-prim-res}
  \text{AUC} = 1 - \left(\frac{\CF\CA}{(\CF+\CA)^2}\right)^n \frac{\Gamma(2n)}{\Gamma(n)\Gamma(1+n)}\,
  \hyp \left(1,2n,1+n;\frac{\CA}{\CF+\CA}\right)\,.
\end{equation}
We can compare the values given by this expression with the ones computed
in~\cite{Larkoski:2019nwj} for the standard definition of \nsub.
\begin{align*}
  n &= 1 &  &\to  & \text{AUC}_{\text{standard}} &= 0.308, & \text{AUC}_{\text{primary}} &= 0.308 \\
  n &= 2 &  &\to  & \text{AUC}_{\text{standard}} &= 0.256, & \text{AUC}_{\text{primary}} &= 0.226 \\
  n &= 3 &  &\to  & \text{AUC}_{\text{standard}} &= 0.231, & \text{AUC}_{\text{primary}} &= 0.173   
\end{align*}
We conclude that, at least at LL, a cut on the primary \nsub $\taup_n$ provides better quark/gluon discrimination power than a cut on the standard \nsub.
The comparison of the two different definitions when evaluated on Monte-Carlo generated pseudo-data will be discussed in section~\ref{sec:montecarlo}.

\subsection{Fixed-coupling limit}\label{sec:fixed-coupling}

In the previous section we have obtained analytic expressions for primary \nsub distributions that are valid at LL, including running-coupling effects. Henceforth, for sake of simplicity, we are going to consider the
fixed coupling limit of Eq.~(\ref{eq:radiator}). In this limit, the probability distributions for quarks and gluons Eq.~(\ref{eq:primdiff-end}) is
\begin{align}\label{eq:primdiff-fc}
  p_i(\taup_1, \dots, \taup_n)   = \left(\frac{\as}{\pi \beta} \right)^n \left(2C_i \right)^n\prod_{j=1}^n\left( \frac{\log \left(1/\taup_j\right)}{\taup_j} \right)\exp\left[-\frac{\as}{ \pi \beta} C_i \log^2\taup_n \right],
  \end{align}
and the likelihood ratio Eq.~(\ref{likelihood-LL}) consequently becomes
\begin{equation}\label{likelihood-LL-fc}
  \mathcal{L}^\text{LL-f.c.}
  =\left( \frac{\CA}{\CF}\right)^n\exp\left[-\frac{\CA-\CF}{\beta}  \frac{\alpha_s}{\pi} \log^2 \taup_n\right].
\end{equation}
In order to further simplify  our notation, we can also reabsorb the
factor $\alpha_s/(\pi \beta)$ in a redefinition of the variables
$\taup_i$, or, equivalently, we can define the colour factors $\CF$
and $\CA$ in units of $\alpha_s/(\pi \beta)$.
We will do the latter, introducing  $\widetilde{C}_i=\frac{\as C_i}{\pi \beta}$.
We will also express Eq.~\eqref{eq:primdiff-fc} in terms of different
variables $x_i=x_i(\taup_i)$.
We therefore rewrite Eq.~(\ref{eq:primdiff-fc}) as
\begin{equation}\label{eq:pgeneric}
  p_i(x_1, \dots, x_n) = r'(x_1) \cdots r'(x_n) \,
  \left(\tC_i^n \, \exp \left[ - \tC_i \, r(x_n) \right] \right)\,.
\end{equation}
In particular, we will consider the following three choices:
\begin{align}
  \text{log square:} && x_i(\taup_i)&=L_i\equiv \log^2\taup_i,
  & r(L_n) & = L_n,
  & r'(L_i) &= 1,\label{eq:plog2}\\
  \text{log:} && x_i(\taup_i)&=l_i\equiv \log(1/\taup_i),
  & r(l_n) &= l_n^2,
  & r'(l_i) &= 2l_i,\label{eq:plog}\\
  \text{linear:} && x_i(\taup_i)&=\taup_i,
  & r(\taup_n)& = \log^2\taup_n,
  & r'(\taup_i)& = 2\frac{\log(1/\taup_i)}{\taup_i}.\label{eq:plin}
\end{align}

\section{Perceptron analytics}\label{sec:analytics}

\subsection{Generic considerations}

\begin{figure}
  \centering
  \begin{subfigure}[t]{0.48\textwidth}
    \centering
    \includegraphics[width=0.6\textwidth]{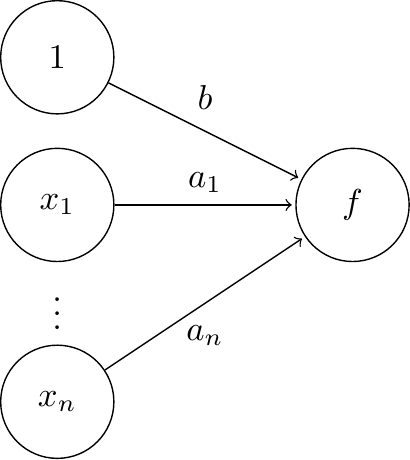}
    \caption{The perceptron consists of $n$ input units and one output
      $y = f(\vec{a} \cdot \vec{x} + b)$.}\label{fig:perceptron-sketch}
  \end{subfigure}
  \hfill
  \begin{subfigure}[t]{0.48\textwidth}
    \centering
    \includegraphics[width=0.75\textwidth]{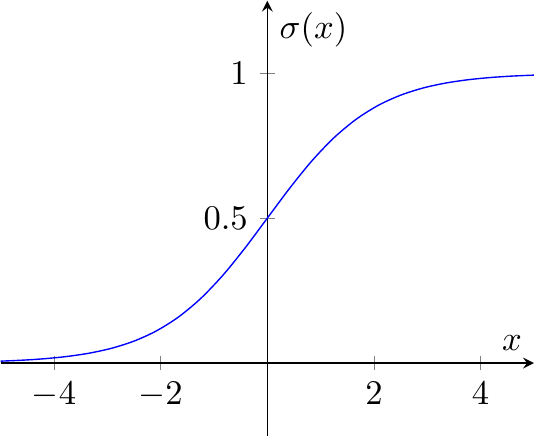}
    \caption{The sigmoid activation function, Eq.~(\ref{eq:sigmoid}).}\label{fig:perceptron-sigmoid}
  \end{subfigure}
  \caption{Sketch of a perceptron (left) and of the sigmoid activation function
    (right).}\label{fig:perceptron}
\end{figure}

In this section we will investigate the simplest neural network, namely the {\em
  perceptron}, where the input layer is directly linked to the output through a
single neuron, as depicted in
Fig.~\ref{fig:perceptron-sketch}. Analytically, this means that the
network output $y$ is a function of
a weighted linear combination of the inputs:
\begin{equation}\label{eq:percfun}
  y= f(\vec{x} \cdot \vec{a} + b) 
\end{equation}
with $\vec{x}$ vector of the input variables, while $\vec{a}$ is the vector of
the {\em weights} of the neuron and $b$ is a {\em bias}. The function $f$,
called {\em activation function}, allows us to introduce an element of non-linearity
in the network. In this paper we will focus on the {\em sigmoid}, which is widely adopted in the ML community. The sigmoid is defined as
\begin{equation}\label{eq:sigmoid}
  f(x) \equiv \sigma(x) = \frac{1}{1+e^{-x}}\,,
\end{equation}
and it is shown in Fig.~\ref{fig:perceptron-sigmoid}.
Alternative functional forms are possible.
For instance, in this context, one could replace the sigmoid with a
rectified version ({\em hard-sigmoid}):
  \begin{equation}
    f(x) = \max(0,\min(x,1))\,.
  \end{equation}

The neural network learns the best choice of the weights by minimising the {\em cost} function, which quantifies the difference between the expected value of the
network $\hat{y}$ and the predicted value $y$ (given by Eq.~(\ref{eq:percfun})). 
In this study we are focusing on the issue of quark versus gluon
discrimination, which is an example of a binary classification
problem, where we have $\hat{y}=0$ (associated with quarks) or
$\hat{y}=1$ (associated with gluons).
In this context, the {\em cross-entropy} loss is one of the most common choice for the loss function
\begin{equation}\label{loss-xse}
  C(y, \hat{y}) = - (1-\hat{y}) \log(1-y) - \hat{y} \log(y)\,.
\end{equation}
This is what we will use for this paper.
However, many of the results we obtain also apply to other loss functions, such as, for instance, the {\em quadratic} loss:
\begin{equation}\label{loss-chi2}
  C(y, \hat{y}) = (y - \hat{y})^2\,.
\end{equation}

In order to train the NN, one usually starts with a so-called training sample, i.e.\ a collection of  input vectors $\{\vec{x}_i\}$, each labelled as a quark jet or as a gluon jet. If we have a training sample of $2 N$ inputs, equally divided between signal and background labels, we can write the cost function as:
\begin{equation}\label{eq:costfundiscr}
  \tC(\vec{a},b) = \frac{1}{2N} \sum_{i= 1}^{N}
  \Big[ C\left( f(\vec{x}_i^{(q)} \cdot \vec{a} + b), 0 \right)
  + C\left( f(\vec{x}_i^{(g)} \cdot \vec{a} + b), 1 \right) \Big] \,.
\end{equation}

The input variables $\vec{x}^{(i)}$ are generated according to a
probability distribution $p_i(\vec{x})$, with $p_q$ ($p_g$) being the
probability distribution of the inputs $\vec{x}$ for quark (gluon)
jets.
If the training sample is large, as it usually is, we can rewrite the above equation in the continuous limit:
\begin{equation}\label{eq:costfun}
  \tC(\vec{a},b) =\frac{1}{2} \int d \vec{x}\, \Big[ p_q(\vec{x})\, C(f(\vec{x} \cdot
    \vec{a} + b),0) + p_g(\vec{x})\, C(f(\vec{x} \cdot \vec{a} + b),1)
    \Big].
\end{equation}
In a general classification problem, the probability distributions of
the inputs are unknown.
However, in the context of QCD studies,
we can exploit expert-knowledge: if we choose the input variables $\vec{x}$ as IRC safe observables, we can apply the powerful machinery of perturbative quantum field theory to determine these distributions at a well-defined and, in principle, systematically improvable accuracy. 

In what follows, we will use the primary \nsub variables $\{\taup_i\}$ as inputs for our perceptron. Using the results obtained in section~\ref{sec:newtau}, we will evaluate Eq.~(\ref{eq:costfun}) using probability distributions calculated at LL accuracy. 
We will then study the global minimum of the cost function.
We will focus on the sigmoid activation function, Eq.~(\ref{eq:sigmoid}), and
the cross-entropy loss, Eq.~(\ref{loss-xse}), although analogous
results can be obtained for the quadratic loss, Eq.~(\ref{loss-chi2}).
More specifically, our goal is to establish whether the set of weights
$\vec{a}$ and bias $b$ that minimises Eq.~(\ref{eq:costfun})
corresponds to a cut on $\taup_n$, which corresponds to the likelihood
ratio at LL accuracy, cf.\ Eq.~(\ref{likelihood-LL}).
This corresponds to checking $a_1=\dots =a_{n-1}=0$.~\footnote{We expect that these coefficients will receive corrections which are suppressed by powers of $\alpha_s$. These corrections are negligible in the strict leading-logarithmic approximation, where only dominants contributions are kept.} Note that the values of $a_n$ and $b$ are not fixed by our LL analysis, even though the network training will converge to some fixed values, corresponding to the minimum of the cost function.
We will show that the ability of the perceptron to find the likelihood ratio crucially depends on the functional form of the input variables. We shall consider three cases, all based on the primary \nsub, namely $\taup_i$, $\log \taup_i$ and $\log^2 \taup_i$.

\subsection{Minimisation of the cost function}

In order to find the extrema of the cost function Eq.~(\ref{eq:costfun}), we
consider the partial derivatives with respect to a generic weight $a_i$ or
$b$. 
With a simple application of the chain rule, we find a set of $n+1$ simultaneous equations
\begin{empheq}[left=\empheqlbrace]{align}
  \frac{\partial{\tC}}{\partial a_i}  &=\frac{1}{2}
  \int d\vec{x} 
  \,  x_i f'(\vec{x} \cdot \vec{a} + b)  
    \Big[ p_q(\vec{x})\,C'(f(\vec{x} \cdot \vec{a} + b),0)
      + p_g(\vec{x})\,C'(f(\vec{x} \cdot \vec{a} + b),1) \Big]
     = 0, \nonumber \\
      \frac{\partial{\tC}}{\partial b}  &=\frac{1}{2}
  \int d\vec{x} \,
 f'(\vec{x} \cdot \vec{a} + b) 
    \Big[ p_q(\vec{x})\,C'(f(\vec{x} \cdot \vec{a} + b),0)
      + p_g(\vec{x})\,C'(f(\vec{x} \cdot \vec{a} + b),1) \Big]
     = 0,
 \nonumber
\end{empheq}
\begin{equation}\label{eq:Cder}
\end{equation}
where $i=1,\dots,n$ and the prime indicates the derivative of a function with respect to its (first) argument.

In general, in order to solve the above  system of simultaneous
equations, we have to explicitly compute the $n$-dimensional integral
in Eq.~(\ref{eq:Cder}). However, in our case, it is possible to find directly a solution at the integrand level. To this purpose, we observe that
 when the equality
\begin{equation}\label{eq:extrcond}
  \frac{p_q(\vec{x})}{p_g(\vec{x})} =
  - \frac{C'(f(\vec{x} \cdot \vec{a} + b),1)}{C'(f(\vec{x} \cdot \vec{a} + b),0)}
\end{equation}
is satisfied for any value of $\vec{x}$, then the system of equations is fulfilled.
Since the cross-entropy loss, Eq.~(\ref{loss-xse}),
satisfies $ C'(y,1) = - \frac{1}{y}$ and $C'(y,0) = - C'(1-y,1)$,
Eq.~(\ref{eq:extrcond}) can be simplified to
\begin{equation}\label{eq:Lsimplify}
  \frac{p_q(\vec{x})}{p_g(\vec{x})} =
  \frac{1- f(\vec{x} \cdot \vec{a} + b)}{f(\vec{x} \cdot \vec{a} + b)}\,.
\end{equation}
We note that this result also holds, for instance, in the case of the quadratic loss.

More specifically, we want to compute Eq.~(\ref{eq:Cder}) for our
probabilities $p_i$ given by Eq.~(\ref{eq:pgeneric}). We
explicitly consider 3 cases differing only by what is used as inputs
to the perceptron: ``log-square inputs'', Eq.~(\ref{eq:plog2}), ``log
inputs'', Eq~(\ref{eq:plog}), and ``linear inputs'', Eq~(\ref{eq:plin}).
In all three cases, we use the sigmoid activation function
and the cross-entropy loss. 
Given the following identities for the sigmoid function:
\begin{equation}
  1-\sigma(x) = \sigma(-x)\,\qquad
  \text{and}\qquad \sigma'(x) = \sigma(x)\,\sigma(-x)\,,
\end{equation}
and the properties of the cross-entropy loss, Eq.~(\ref{eq:Cder}) may be rewritten as:
\begin{empheq}[left=\empheqlbrace]{align}
%\begin{align}
\label{eq:Cder-sig-xe}
  \frac{\partial{\tC}}{\partial a_i}  &=\frac{1}{2}
  \int d\vec{x} 
  \,  x_i \, r'(x_1) \cdots r'(x_n) \nonumber \\
  & \qquad\qquad \Big[ \tCF^n \, e^{- {\tCF} \, r(x_n)}
    \, \sigma(\vec{x} \cdot \vec{a} + b)
    - \tCA^n \, e^{- \tCA \, r(x_n)}
    \, \sigma(-\vec{x} \cdot \vec{a} - b) \Big]
     = 0, \nonumber \\
  \frac{\partial{\tC}}{\partial b}  &=\frac{1}{2}
  \int d\vec{x} 
  \,  r'(x_1) \cdots r'(x_n) \nonumber \\
  & \qquad\qquad \Big[ \tCF^n \, e^{- \tCF \, r(x_n)}
    \, \sigma(\vec{x} \cdot \vec{a} + b)
    - \tCA^n \, e^{- \tCA \, r(x_n)}
    \, \sigma(-\vec{x} \cdot \vec{a} - b) \Big]
     = 0.
%\end{align}
\end{empheq}

\paragraph{Log-square inputs.}
Let us start by considering $\logtsq_i= \log^2 \taup_i$ as inputs to the perceptron. In this case the probability distributions for quarks and gluons in the fixed-coupling limit are given by Eq.~(\ref{eq:plog2}). This is a very lucky scenario
because we can determine the minimum at the integrand level.
Indeed, the condition Eq.~(\ref{eq:Lsimplify}) gives the constraint
\begin{equation}\label{eq:sigmoidlogsquare}
  \left(\frac{\tCF}{\tCA}\right)^n \, \exp\left[- (\tCF-\tCA) \logtsq_n\right]
  = \exp\left[-\vec{a} \cdot \vec{\logtsq} - b\right]\,,
\end{equation}
leading to the following solution
\begin{equation}\label{eq:minlogsq}
  a_1 = \dots = a_{n-1} = 0\,, \qquad a_n = \tCF - \tCA, \qquad
  b = n \log\left(\frac{\tCA}{\tCF}\right)\,.
\end{equation}
Hence, for the log-square inputs, the minimum of the cost function
does agree with
the optimal cut on $\taup_n$ dictated by the likelihood.

Note that, as $\tCF < \tCA$, the weight $a_n$ is negative.
This is expected, since the sigmoid function is monotonic and
we have mapped the gluon (quark) sample to output 1 (0), see Eq.~(\ref{eq:costfundiscr}),
whereas the gluon sample has larger $\taup_i$ and thus smaller $\logtsq_i$:
the negative sign of $a_n$ restores the proper ordering between inputs and output of the perceptron.

If we restrict ourselves to the case $n=2$, it is also possible to explicitly perform the integrals in Eq.~(\ref{eq:costfun}) and arrive at an analytic expression for the cost function. We report this calculation in appendix~\ref{app:cost-function}.

\begin{figure}
  \begin{subfigure}[t]{0.48\textwidth}
    \centering
    \includegraphics[width=\textwidth,page=5]{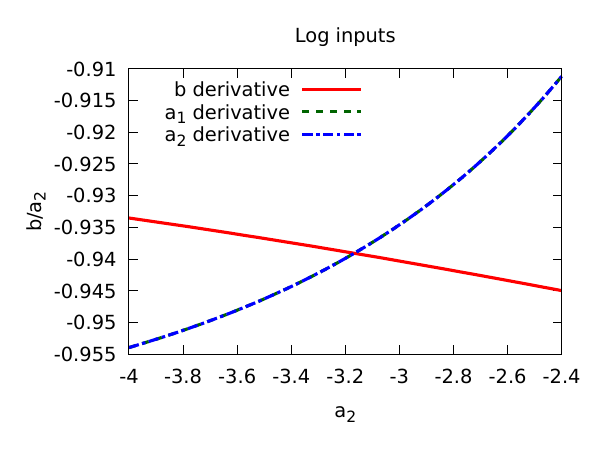}
    \caption{Log inputs}\label{fig:num-solutions-log}
  \end{subfigure}
  \begin{subfigure}[t]{0.48\textwidth}
    \centering
    \includegraphics[width=\textwidth,page=6]{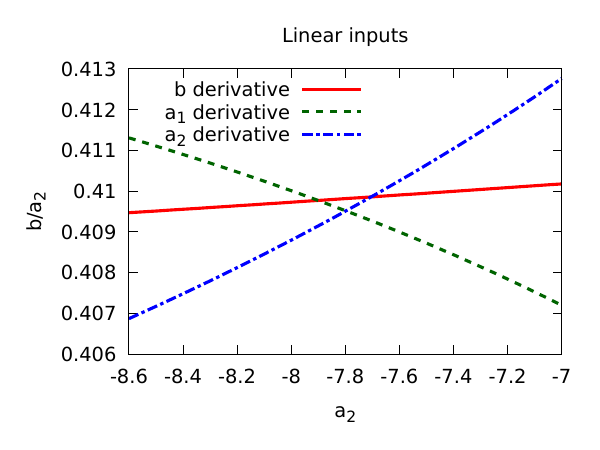}
    \caption{Linear inputs}\label{fig:num-solutions-lin}
  \end{subfigure}
  \caption{Solutions of Eq.~(\ref{eq:Cder-sig-xe}), in the case $n=2$ and with $a_1=0$, plotted as points in the $(a_2,b/a_2)$ plane,
  for log (left) and linear (right) inputs.}\label{fig:num-solutions}
\end{figure}

\paragraph{Log inputs.}
We now turn our attention to logarithmic inputs $\logt=-\log \taup_i$.
In this case we are not able to determine the position of the minimum
at the integrand level and we are forced to use the explicit
constraints from Eq.~(\ref{eq:Cder-sig-xe}).
In particular, we want to check if the likelihood condition $a_1 = \dots = a_{n-1} = 0$
is still a solution of the system of this system of $n+1$ equations. To this  purpose,
we use Eq.~(\ref{eq:Cder-sig-xe}) with the probability distribution given by Eq.~(\ref{eq:plog}),
then we set $a_1 = \dots = a_{n-1} = 0$, thus explicitly obtaining
\begin{empheq}[left=\empheqlbrace]{align}
%\begin{align}
\label{eq:Cderlog}
  \frac{\partial{\tC}}{\partial a_i}  &=
  2^{n-1} \int_0^{\infty} d\logt_n\,\logt_n\,I_i(\logt_n)
  \left[
    \frac{\tCF^n \exp(-\tCF \logt_n^2)}{1 + \exp(-a_n \logt_n - b)}
    - \frac{\tCA^n \exp(-\tCA \logt_n^2)}{1 + \exp(a_n  \logt_n + b)}
  \right] = 0, \nonumber \\
    \frac{\partial{\tC}}{\partial b}  &=
  2^{n-1} \int_0^{\infty} d\logt_n\,\logt_n\,I_0(\logt_n)
  \left[
    \frac{\tCF^n \exp(-\tCF \logt_n^2)}{1 + \exp(-a_n \logt_n - b)}
    - \frac{\tCA^n \exp(-\tCA \logt_n^2)}{1 + \exp(a_n  \logt_n + b)}
  \right] = 0,
%\end{align}
\end{empheq}
where $I_i(\logt_n)$ is the result of the integration over $\logt_1, \dots,\logt_{n-1}$
\begin{equation}\label{eq:Ilogt}
  I_i(\logt_n) =
  \int_{0}^{\logt_n} d\logt_{n-1}\,\logt_{n-1} \dots
  \int_{0}^{\logt_2}d\logt_1\,\logt_1\,v_i\,, \quad \text{with} \quad v_0 = 1\,,\; v_i = \logt_i\,.
\end{equation}
Up to an irrelevant overall multiplicative constant, these integrals
are easily found to be
\begin{equation}
  I_0(\logt_n) \propto \logt_n^{2n-2}\,, \quad
  I_i(\logt_n) \propto \logt_n^{2n-1} \quad \text{for } \quad i = 1, \dots, n\,.
\end{equation}
Replacing this result in Eq.~(\ref{eq:Cderlog}), we see that all of the
derivatives with respect to $a_i$, $i = 1, \dots, n$ give rise to the same
equation, and thus the system of $n+1$ simultaneous equation reduces to a system of just two independent equations, for any $n$. These two equations correspond to two lines in the $(a_n,b)$ plane. If these two lines never cross, then the system has no solution, the minimum of the cost function is not at $a_1=\dots a_{n-1}=0$ and thus the perceptron is not able to correctly reproduce the likelihood. If instead these lines meet at some $(\bar{a}_n, \bar{b})$, then the minimum of the cost function does correspond to the likelihood ratio. Despite numerous attempts, we have not been able to perform the final integration over $l_n$ in Eq.~(\ref{eq:Cderlog}) analytically. However, we can perform the integration numerically and plot the result in the $(a_n,b)$ plane. This is done in Fig.~\ref{fig:num-solutions-log}, where, without loss of generality, we have concentrated on the case $n=2$. It is clear from the plot that the two curves do meet in a point and hence the perceptron is able to find the correct minimum, i.e.\ the one dictated by the likelihood with $a_1=0$.

The explicit $n$-dependence of the coefficients $a_n$ and $b_n$ can be found numerically by solving the system of equations.  Furthermore, it is possible to obtain analytically the scaling of $a_n$ with respect the colour factors, as detailed in appendix~\ref{app:scaling}. We find that, once we have factored out $\tCF^{-1/2}$, the resulting coefficient only depends on the ratio of colour factors:
\begin{equation}
  \frac{a_n}{\sqrt{\tCF}} = F \left(\frac{\tCA}{\tCF}\right),
\end{equation}
where $F$ is a function that we have not determined.

\paragraph{Linear inputs.}
We now move to consider linear inputs of the perceptron, i.e.\ the
variables $\taup_i$ directly. We follow the same logic as in the case
of the logarithmic inputs, namely we want to check whether there is a
minimum of the cost function satisfying $a_1=\dots =a_{n-1}=0$.
Following the same steps as before, we have
\begin{empheq}[left=\empheqlbrace]{align}
%\begin{align}
  \frac{\partial{\tC}}{\partial a_i}  &=
  2^{n-1} \int_0^{1} \frac{d \taup_n}{\taup_n}\,\log\frac{1}{\taup_n}\,I_i(\taup_n) \left[
    \frac{\tCF^n \exp(-\tCF \log^2\taup_n)}{1 + \exp(-a_n \taup_n - b)}
    - \frac{\tCA^n \exp(-\tCA \log^2\taup_n)}{1 + \exp(a_n \taup_n + b)}
  \right] = 0, \nonumber \\
    \frac{\partial{\tC}}{\partial b} &=
  2^{n-1} \int_0^{1} \frac{d \taup_n}{\taup_n}\,\log\frac{1}{\taup_n}\,I_0(\taup_n) \left[
    \frac{\tCF^n \exp(-\tCF \log^2\taup_n)}{1 + \exp(-a_n \taup_n - b)}
    - \frac{\tCA^n \exp(-\tCA \log^2\taup_n)}{1 + \exp(a_n \taup_n + b)}
  \right] = 0,\nonumber
%\end{align}
\end{empheq}
\begin{equation}\label{eq:Cderlinear}
\end{equation}
with \begin{equation}\label{eq:Iinear}
  I_i(\taup_n) =
  \int_{\taup_n}^{1} \frac{d \taup_{n-1}}{\taup_{n-1}}\,\log\frac{1}{\taup_{n-1}} \dots
  \int_{\taup_2}^{1} \frac{d \taup_{1}}{\taup_{1}}\,\log\frac{1}{\taup_{1}}\,w_i\,, \quad \text{with} \quad w_0 = 1\,,\; w_i = \taup_i\,. 
\end{equation}
There is a crucial difference between the integrals in
Eq.~(\ref{eq:Iinear}) and the corresponding ones in the case of the
logarithmic inputs, Eq.~(\ref{eq:Ilogt}): the cases
with $1\le i\le n$ lead to $n$ different constraints. 
For sake of simplicity, let us consider $n=2$. We have
\begin{equation}
  I_i(\taup_2) =
  \int_{\taup_2}^1 \frac{d\taup_1}{\taup_1}\, \log \frac{1}{\taup_1}\,v_i =
  \begin{cases}
    \frac{1}{2} \log^2 \taup_2\,,        & i = 0, \\
    1 - \taup_2 + \taup_2 \log \taup_2\,, & i = 1,  \\
    \frac{1}{2} \taup_2 \log^2 \taup_2\,, & i = 2 .
  \end{cases}  
\end{equation}
Thus, all the three equations appearing in (\ref{eq:Cderlinear}),
i.e.\ $i=0,1,2$ provide independent conditions. The system has solutions if the corresponding three curves in the $(a_2,b)$ plane meet in one point. By numerically performing the integrations, we can check whether this happens or not. 
This is done in Fig.~\ref{fig:num-solutions-lin} where it is clear
that the three curves do not intersect at a common point and hence the
perceptron is unable to find the correct minimum dictated by the likelihood.

Let us summarise the findings of this section. Working in the leading
logarithmic approximation of QCD, we have analytically studied the
behaviour of a perceptron with sigmoid activation and cross-entropy
cost function, in the context of a binary classification problem.
We have explicitly considered three variants of the primary \nsub inputs: squared logarithms, logarithms and linear inputs. 
In the first two cases the minimum of the cost function captures the
expected behaviour from the likelihood ratio, i.e.
$a_1=\dots =a_{n-1}=0$.
This does not happen with linear inputs, although the configuration
$a_1=\dots =a_{n-1}=0$ is within the reach of the network.
This is most likely due to the fact that the simple perceptron with
linear inputs struggles at correctly learning the probability
distributions which are intrinsically logarithmic.
We expect that a more complex network would be needed in this case and
this is shown explicitly in section~\ref{sec:montecarlo}.

\section{Perceptron numerics}\label{sec:numerics}

\begin{figure}
  \centering
  \begin{subfigure}[t]{0.32\textwidth}
    \includegraphics[width=\textwidth, page=1]{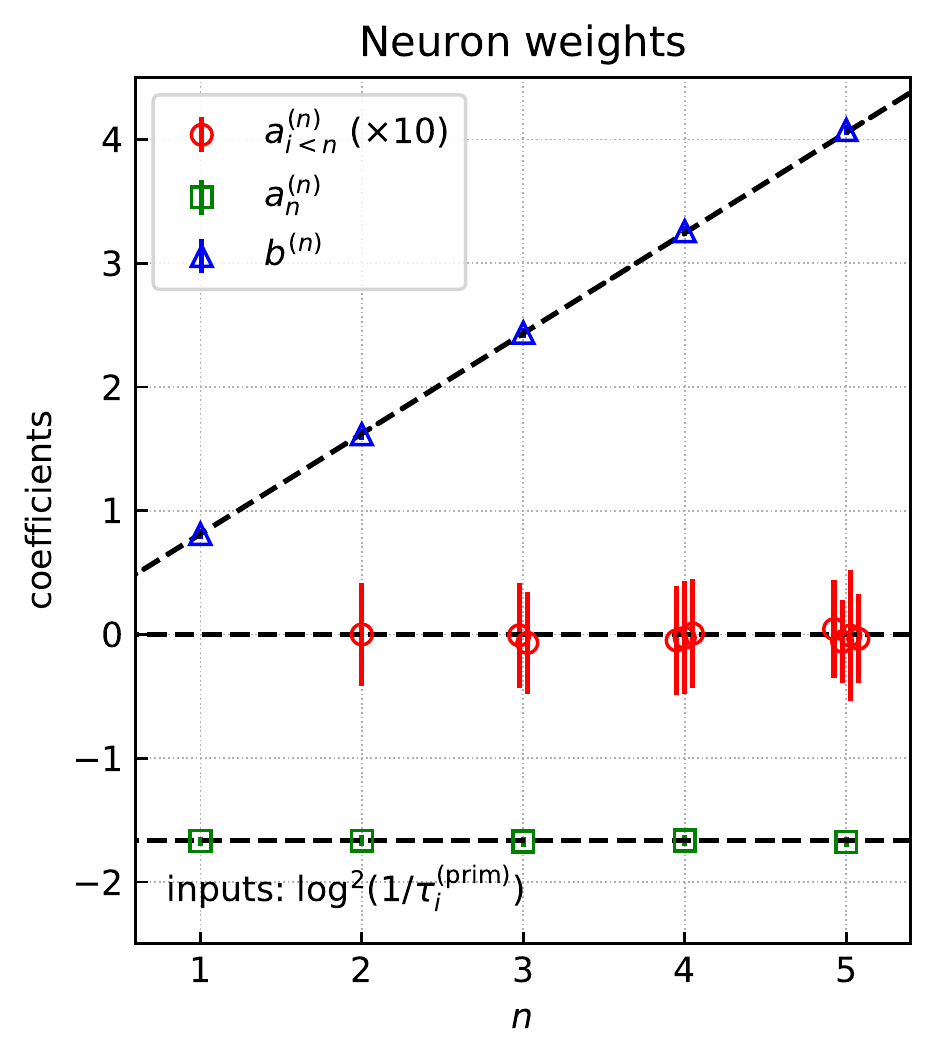}
    \caption{Inputs: $\log^2\taup_i$}
  \end{subfigure}
  \hfill
  \begin{subfigure}[t]{0.32\textwidth}
    \includegraphics[width=\textwidth, page=2]{figures/perceptron-params-v-analytic.pdf}
    \caption{Inputs: $\log\taup_i$}
  \end{subfigure}
  \hfill
  \begin{subfigure}[t]{0.32\textwidth}
    \includegraphics[width=\textwidth, page=3]{figures/perceptron-params-v-analytic.pdf}
    \caption{Inputs: $\taup_i$}
  \end{subfigure}
  \caption{Perceptron parameters after training. When available,
    expected analytic results are shown as dashed lines.}  \label{fig:coeffs}
\end{figure}

In this section we validate our analytic findings with an actual
implementation of a perceptron.
In practice, we have used a simple implementation based
on~\cite{NNlectures}, with a learning rate of $0.1$ and a mini-batch
size of~32.\footnote{We have also cross-checked our results against a
  standard PyTorch implementation.
  % 
%  Conversely, the full NN presented in Section~\ref{sec:montecarlo}
%  below have been checked to give results comparable with our
%  implementation based on Ref.~\cite{NNlectures} with a fully
%  connected network with 3 layers of 64 (sigmoid) neurons each and
%  gradient descent.
  }
Because our first-principle analysis has been developed at LL accuracy, in order to numerically test the perceptron performance we generate a sample of pseudo-data according to the QCD LL distribution for quark and gluon jets. 
We consider the three different input variants also used in the analytic study, namely square logarithms, logarithms and the linear version of the \nsub inputs. 
In order to train our single-neuron network we use a sample of 1M events in the first two cases and 16M the for linear inputs, unless otherwise stated.
Furthermore, we perform our study as a function of number of \nsub
variables. Specifically, when we quote a given value of $n$, we imply
that all $\taup_i$ with $i\le n$ have been used as inputs.
The results of this study are collected in Fig.~\ref{fig:coeffs}. Each
plot shows the value of the network weights  $a_i$ and $b$ after
training, i.e.\ at the minimum of the cost function that has been
found by the network through back-propagation and gradient descent.
The plot on the left is for log-square inputs, the one in the middle
for log inputs and the one on the right for linear inputs. The values
of the weights determined by the network are shown as circles (for
$a_i$ with $i<n$), squared (for $a_n$) and triangles (for $b$), with
the respective numerical uncertainties. In the case of log-square and
log inputs, we also show the behaviour obtained from our analytic
calculations in section~\ref{sec:analytics}. We find perfect
agreement. In particular, the minimum of the cost function found by
the network is consistent with $a_1=\cdots= a_{n-1}=0$. This does not happen in the case of linear inputs, although the discrepancy is tiny.

\begin{figure}
  \centering
  \begin{subfigure}[t]{0.48\textwidth}
    \includegraphics[width=\textwidth, page=3]{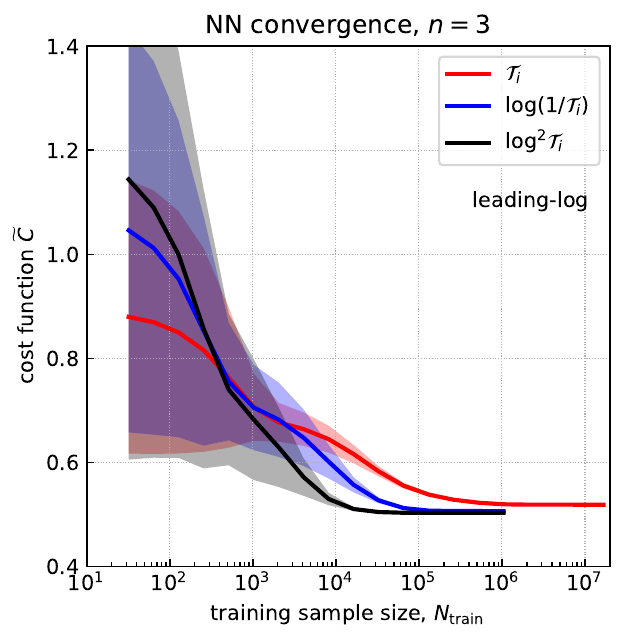}
    \caption{Inputs: $n=3$}
  \end{subfigure}
  \hfill
  \begin{subfigure}[t]{0.48\textwidth}
    \includegraphics[width=\textwidth, page=5]{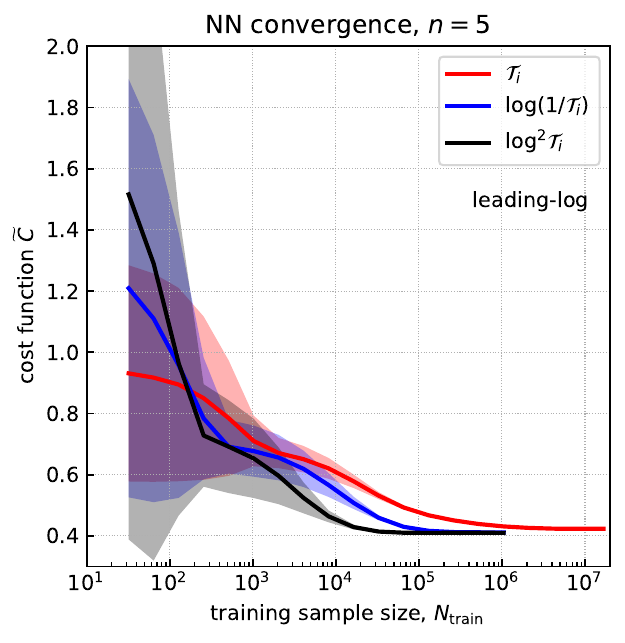}
    \caption{Inputs: $n=5$}
  \end{subfigure}
  \caption{Convergence of the network as a function of the training
    sample size.}  \label{fig:convergence} 
\end{figure}

\begin{figure}
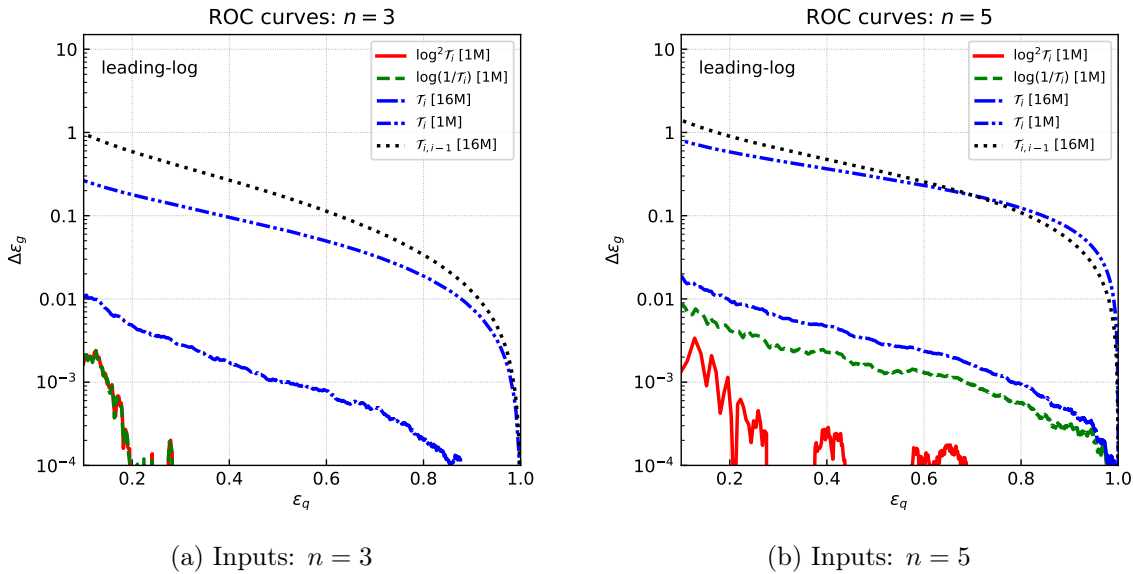

  \centering
  \begin{subfigure}[t]{0.48\textwidth}
    \includegraphics[width=\textwidth, page=7]{figures/perceptron-rocs.pdf}
    \caption{Inputs: $n=3$}
  \end{subfigure}
  \hfill
  \begin{subfigure}[t]{0.48\textwidth}
    \includegraphics[width=\textwidth, page=9]{figures/perceptron-rocs.pdf}
    \caption{Inputs: $n=5$}
  \end{subfigure}
  \caption{ROC curves obtained from a trained perceptron with
    different inputs. Note that, in order to highlight deviations from optimal performance, we show Eq.~(\ref{eq:DELTA_ROC}). Thus, the closer these curves are to zero, the better the performance.}  \label{fig:perceptron-roc}
\end{figure}

It is interesting to investigate whether the theoretical issues for
the linear inputs
have some visible effect on the network performance. In order to do so, we first perform a study of the perceptron convergence as a function of the training sample size $N_\text{train}$.
As before, we repeat this study for each of the input variants
previously introduced. We also consider two different values of $n$: for the case $n=3$ we build our inputs from $\taup_1,\taup_2, \taup_3$, while for $n=5$, we also include $\taup_4$ and $\taup_5$.
In Fig.~\ref{fig:convergence} we plot the cost function as a function of the training sample size.
Fig.~\ref{fig:convergence} is obtained by training the perceptron with a
progressively increasing sample of pseudo-data generated on the fly according to
the QCD LL distribution. At fixed values of $N_\text{train}$, the cost function
$\tC$ of the trained network is then evaluated on a test sample of fixed
dimension.
This procedure is iterated a number of times, and at the end, for each
$N_\text{train}$, we take the average and standard deviation of the
different runs as a measure of the central value and uncertainty.
The plots clearly show that the convergence with linear inputs is slower than in the cases of log-squares and logs, exposing the fact that the single-neuron network struggles to learn intrinsically logarithmic distributions with linear inputs. 

Furthermore, with the same set up, we can study the actual performance
of the network using ROC curves. To highlight deviations from the
ideal case dictated by the likelihood ratio, the plots in Fig.~\ref{fig:convergence} show
\begin{equation}\label{eq:DELTA_ROC}
\Delta \epsilon_g=\frac{\text{ROC}_{NN}}{\text{ROC}_\text{lik.}}-1,
\end{equation}
where $\text{ROC}_{NN}$ is the network ROC curve, while $\text{ROC}_\text{lik.}$ is the ROC curve that corresponds to a cut on $\taup_n$. 
We perform this study for two different sizes of the training sample:
1M and 16M. We know from Fig.~\ref{fig:convergence} that the former is
enough to achieve convergence in the case of log-square and log
inputs but not for linear inputs. Thus, we expect to see larger
differences in this case. This is indeed confirmed by the plots in
Fig.~\ref{fig:perceptron-roc}. The dash-dot-dotted curve in blue,
corresponding to linear inputs and a training done on a sample of 1M
events, indicates that the difference with respect the optimal case is
of order 10\% for quark efficiencies $\epsilon_q<0.6$ in the case of $n=3$  and exceeds
30\% in the same region for $n=5$. If the training sample size is
increased to 16M, the perceptron with linear inputs still performs
worse than the other choices, although the difference after training
is now small and never exceeds 1\%. 
For comparison, we also show the results for \nsub ratios as
inputs. In this case, for $n=3$, we include $\taup_1$,
$\taup_{21}=\frac{\taup_2}{\taup_1}$ and
$\taup_{32}=\frac{\taup_3}{\taup_2}$.
In this case, the expected ideal cut on $\taup_n$ cannot be reproduced
by any choice of the perceptron weights $a_i$ and bias $b$. 
The perceptron performance in this case is consequently much worse
even after a training on our largest sample.

\section{Monte Carlo studies}\label{sec:montecarlo}

In the previous section, we have validated the results of
section~\ref{sec:analytics}, using the same
setting as for our analytical calculations, namely a single-neuron
network fed with primary \nsub variables (or functions thereof)
distributed according to QCD predictions at leading-logarithmic accuracy.
This setup is somehow oversimplified, and one may wonder how our results compare
in term of performance and convergence to a fully-fledged neural network trained
on Monte Carlo pseudo-data.
In addition, we have adopted the primary \nsub definition, whose nice analytical
properties naturally justifies its use in our theoretical studies. However,
this alternative definition so far has been compared to the standard definition
only in terms of AUC at LL accuracy. Even though the purpose of this work is not
a detailed comparison of the two definitions, we need to make sure that primary \nsub performs sensibly as a
quark/gluon tagger.

The purpose of this section is to address some of these concerns, by extending
the setup of section~\ref{sec:numerics} to a more realistic scenario, both in terms of the generation of pseudo-data and of the network architecture employed in the analysis. 
First, we generate pseudo-data with
Pythia~8.230~\cite{Sjostrand:2014zea}, including hadronisation and
multi-parton interactions. We simulate dijet events
$qq \to qq$ and $gg \to gg$ and we cluster jets with the anti-$k_t$
algorithm~\cite{Cacciari:2008gp}, as implemented in
FastJet~\cite{Cacciari:2011ma}, with jet radius $R_0 = 0.4$. We keep
jets with $p_t > 2$~TeV and $|y|<4$. We then compute $\tau_N$ and
$\taup_N$ (i.e.\ both \nsub definitions) up to the desired $N=n$ for
each jet. We set the \nsub parameter $\beta$ to 1.
For the standard \nsub, we use the reference axes obtained from the
exclusive $k_t$ axis with winner-takes-all~\cite{Larkoski:2014uqa} recombination
and a one-pass minimisation~\cite{Thaler:2011gf}.
In addition to linear, log-square and log functional forms and \nsub
ratios already adopted in the previous section, we also consider a
running-coupling-like input, obtained by evaluating the
radiator $\mathcal{R}(\tau_i)$ of Eq.~(\ref{eq:radiator}) with a
one-loop running coupling.\footnote{We set
  $\alpha_s(p_tR_0)=0.09$, and we regulate the divergence at the Landau
  pole by freezing the coupling at a scale $k_t$ such that
  $2\alpha_s(p_tR)\beta_0\log(p_tR_0/k_t)=0.99$.}

The samples thus obtained are used to train either a single perceptron
(as in the previous section) or a full neural network.
The perceptron is trained using the same implementation
based on Ref.~\cite{NNlectures}, training over 50 epochs, with a learning rate of 0.5.
The \textit{full network} uses a fully connected architecture implemented
in PyTorch~1.4.0. The number of inputs equals the number $n$ of input
\nsub values (between 2 and 7). There are three hidden layers with 
128~nodes per layer and a LeakyReLu~\cite{leaky} activation function with negative slope of 0.01 
is applied following each of these layers. We train the network using binary cross-entropy
for a maximum of 50 epochs using the Adam~\cite{adam} optimiser with a learning
rate of~0.005 and a mini-batch size of 32 events. During training, the 
area under curve is evaluated after each epoch on a statistically independent validation
set and training terminates if there is no improvement
for 5 consecutive epochs~\cite{early}. The best model on the validation set is then selected
and evaluated on another independent test set for which results are reported. 
Data are split 6:2:2 between training, validation, and testing. No extensive optimisation
of hyperparameters has been performed and training usually terminates due to the early stopping
criterion before reaching the maximally possible number of epochs.

\begin{figure}
  \centering
  \begin{subfigure}[t]{0.48\textwidth}
    \includegraphics[width=\textwidth, page=1]{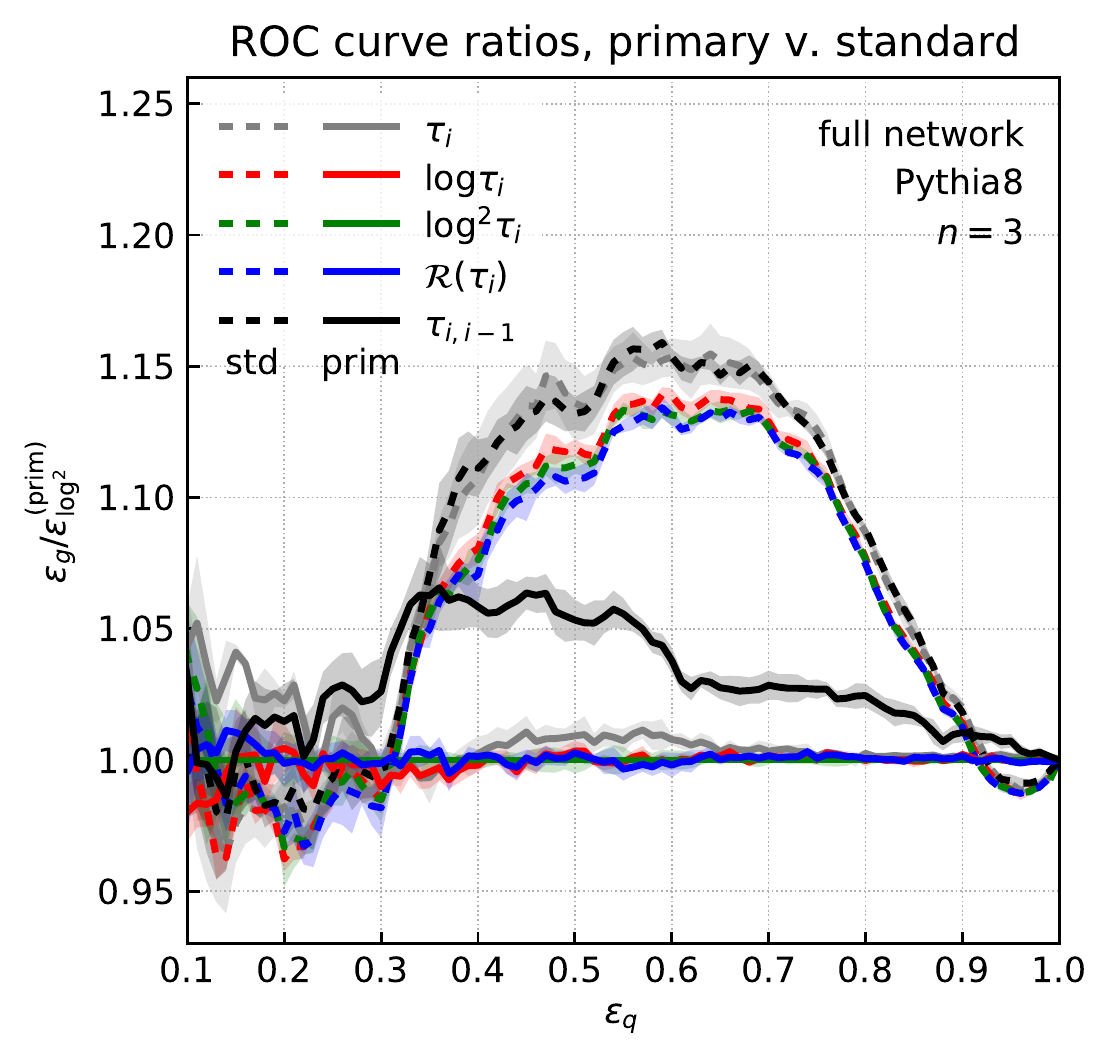}
    \caption{Performance comparison between the standard
      $N$-subjettiness (dashed lines) and the primary $N$-subjettiness
      (solid lines). All results are obtained using a full neural network.}
    \label{fig:rocPRIvsFUL}
  \end{subfigure}
  \hfill
  \begin{subfigure}[t]{0.48\textwidth}
    \includegraphics[width=\textwidth, page=2]{figures/pythia-rocs.pdf}
    \caption{Performance comparison between a single perceptron
      (dashed lines) and a full neural network (solid lines). All
      results are obtained using primary $N$-subjettiness.}
    \label{fig:rocPERvsNET}    
  \end{subfigure}
  \caption{ROC curves obtained after training on a Pythia8 sample,
    using the \nsub variables, with $n=3$.
    To better highlight differences the vertical axes show the gluon
    rate normalised by what is obtained using $\log^2\taup_i$
    inputs for primary \nsub trained on a full neural
    network. 
    The different colours correspond to different inputs to the neural
    network.}\label{fig:roc-pythia}
\end{figure}

We start with a comparison of the primary and the standard definition of \nsub
in terms of network performance. In Fig.~\ref{fig:rocPRIvsFUL} we plot the ROC
curves for the primary (solid lines) and standard (dashed line)
definitions of \nsub, and for different choices of the input functional form (different
colours). The gluon efficiencies are normalised by the central value
obtained with primary \nsub with $\log^2(\taup_i)$ inputs
trained on a full NN.
The central values and the error bands are calculated by taking the
average and standard deviation of five different runs with randomly initialised weights.

We see that the performance of networks trained with the standard definition is worse by
10-15\% compared to the primary definition for mid-values of the quark
efficiency, $0.4 \lesssim\varepsilon_q \lesssim 0.8$ and comparable
elsewhere, except for a small region at large quark efficiency, $\varepsilon_q
 \gtrsim 0.9$. 
Even though the benefit of using primary \nsub is not as
large as one could have expected based on our leading-logarithmic
calculations, we still observe a performance improvement.
We note also that at very small quark efficiency non-perturbative effects
have a sizeable effect, invalidating our arguments purely based on a
perturbative calculation.
Furthermore, large $\varepsilon_q$ correspond to the regime where the
cut on \nsub is no longer small and our arguments based on
the resummation of large logarithms no longer apply, and one should
instead consider the full structure of hard matrix
elements.\footnote{Note that in that region, our Monte-Carlo
  pseudo-data, based on Pythia8, does not contain all the details of
  the physics in any case.}
That said, Fig.~\ref{fig:rocPRIvsFUL} shows that for
the most of the range in quark efficiency, a better discrimination is
reached if we adopt the primary-\nsub definition.

It is also interesting to compare the results obtained with different
inputs in Fig.~\ref{fig:rocPRIvsFUL}. Although one should expect that
a full NN should ultimately reach the performance of the likelihood independently
on the choice of inputs, our practical setup still shows a mild
dependence on the choice of inputs.
The key point is that the favoured inputs agree with our analytic
expectations from section~\ref{sec:analytics} with logarithmic inputs
($\log\tau_i$, $\log^2\tau_i$ and $\mathcal{R}(\tau_i)$) showing an
equally-good performance, the linear inputs only marginally worse and
the ratio inputs showing a (few percent) worse discriminating power.
\footnote{The choice of logarithmic inputs was also found to be helpful, on a phenomenological basis, in Ref.~\cite{Dreyer:2018nbf}.}
Even though the convergence of the neural network
is more delicate for the case of ratio inputs (and, to a lesser
extent, for linear inputs), a performance similar to the optimal one
reported in Fig.~\ref{fig:rocPRIvsFUL} for logarithmic inputs could be
achieved with a careful optimisation of the hyperparameters.

\begin{figure}
  \centering
  \begin{subfigure}[t]{0.32\textwidth}
    \includegraphics[width=\textwidth, page=1]{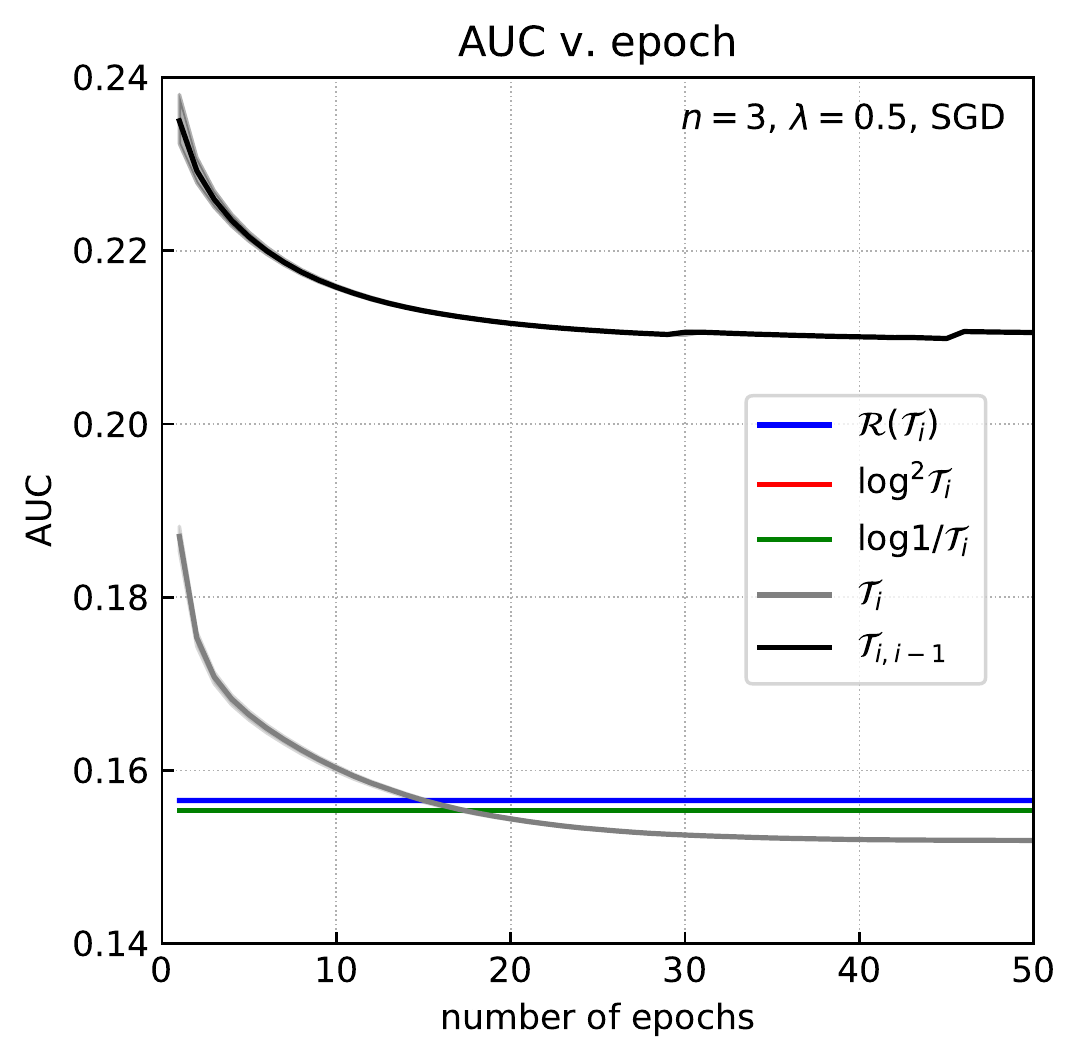}
    \caption{AUC as a function of the number of epochs for a perceptron.}
    \label{fig:auc-v-epoch}
  \end{subfigure}
  \hfill
  \begin{subfigure}[t]{0.32\textwidth}
    \includegraphics[width=\textwidth, page=1]{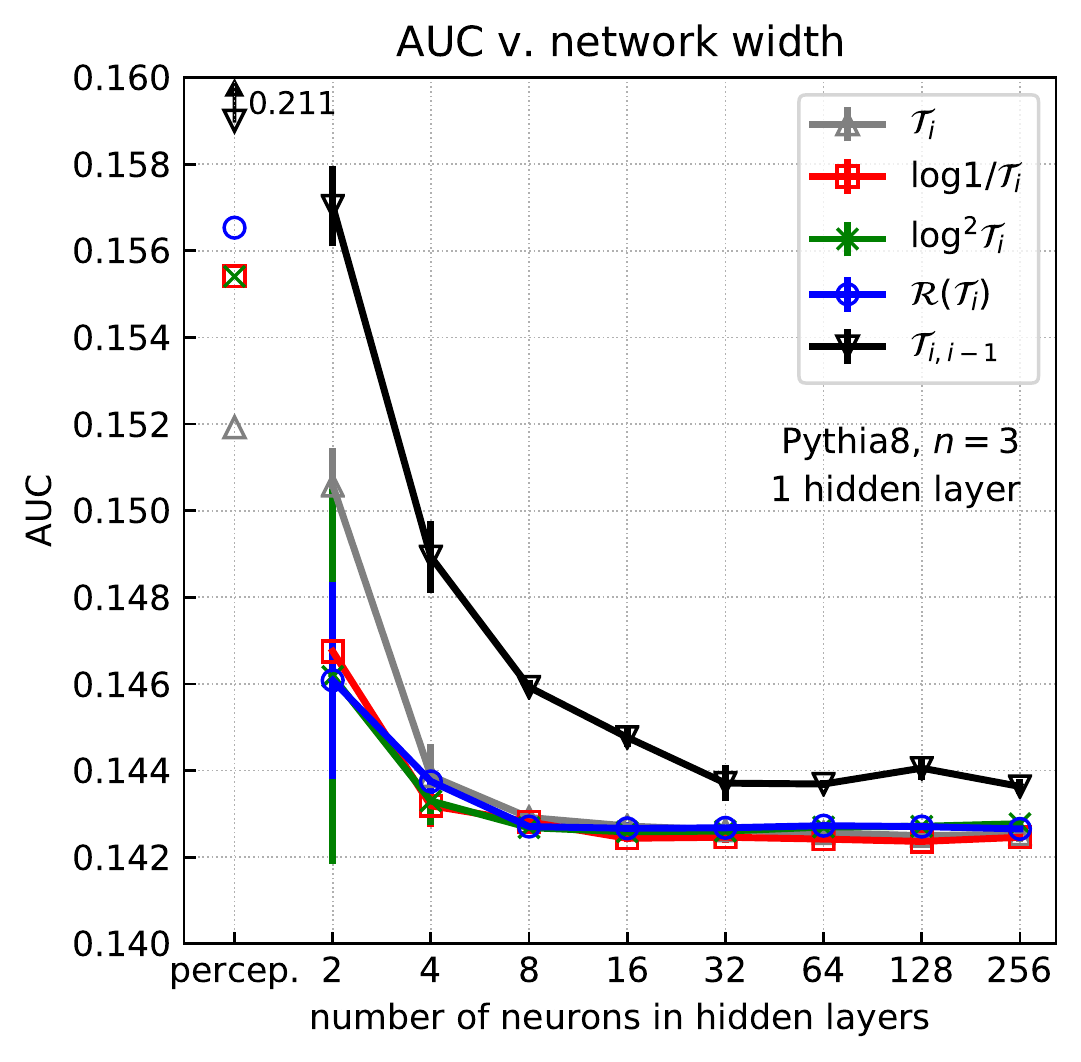}
    \caption{AUC as a function of the number of neurons in a single
      hidden layer.}
    \label{fig:auc-v-width}    
  \end{subfigure}
  \hfill
  \begin{subfigure}[t]{0.32\textwidth}
    \includegraphics[width=\textwidth, page=2]{figures/AUC-v-netsize.pdf}
    \caption{AUC as a function of the number of hidden layers (with 128
      neurons per layer).}
    \label{fig:auc-v-depth}    
  \end{subfigure}
  \caption{AUC convergence study for the Pythia8 sample as a function
    of the network parameters: number of training epochs, network
    width and network depth. The different inputs are represented with
    the same colours as in Fig.~\ref{fig:roc-pythia}.}\label{fig:auc-pythia}
\end{figure}

We now move to Fig.~\ref{fig:rocPERvsNET}, where we compare the ROC
curves obtained with a full neural network to those of a simple
perceptron. In this plot we select the primary \nsub definition and we
display results for the usual input choices.
The solid lines in Fig.~\ref{fig:rocPERvsNET} coincide with the ones
in Fig.~\ref{fig:rocPRIvsFUL}.
First, we observe that a perceptron trained with Monte Carlo pseudo-data
performs worse compared to the full NN for all the considered input
types.
This is not surprising as the arguments in section~\ref{sec:analytics}
are based on a simplified, leading-logarithmic, approach and subleading
corrections can be expected to come with additional complexity that a
single perceptron would fail to capture.
It is actually remarkable that for $\varepsilon_q \gtrsim 0.6$
a single perceptron
typically gives performances which are only 10\%
worse than the full network.
This is not true if we consider \nsub ratio inputs, in which case  the perceptron performance
is sizeably worse than
the performance we get by using the full NN. This in agreement with the
observation made towards the end of section~\ref{sec:numerics}, namely
that a more complex network architecture is able to learn the 
correct weights for the ratio inputs, while the perceptron is not. 

The behaviour of the other input types is relatively similar,
although, quite surprisingly, linear inputs tend to give a slightly better
performance than logarithmic ones. 
The origin of this is not totally
clear. One can argue that the difference in performance between linear
and logarithmic inputs was small at leading-logarithmic accuracy and
that several effects --- subleading logarithmic effects, hard
matrix-element corrections, non-perturbative effects, ... --- can lift
this approximate degeneracy.
 One should also keep in mind that our
leading-logarithmic calculation is valid at small-enough quark/gluon
efficiencies, while, as the efficiency grows, we become sensitive to contributions from other
kinematic domains.
We also note that if we increase the jet $p_t$, e.g.\ if we look at
20~TeV jets at 100~TeV $pp$ collider, the difference between
logarithmic and linear inputs becomes again comparable, which is likely
a sign that the phase-space at larger $p_t$ has an increased
contribution from the resummation region.  
We finally note that a simple cut on $\taup_n$, which is optimal at LL, gives comparable results to the perceptron at large quark efficiencies $\varepsilon_q \gtrsim 0.7$, but performs rather worse otherwise.

Next, we study how the convergence of the network is affected by the
choice of inputs.
In this context, convergence can mean two different things. Firstly,
we can speak about convergence of the network training (as we did in
Fig.~\ref{fig:convergence}). This is shown in
Fig.~\ref{fig:auc-v-epoch}, for the  area under the ROC curve (AUC) as a function of the number of
training epochs for a single perceptron.
For this plot, we have used primary \nsub and (mini-batch) stochastic
gradient descend with a learning rate $\lambda=0.5$.
The better performance obtained for linear inputs is confirmed in
Fig.~\ref{fig:auc-v-epoch}. However, this happens at a cost in
convergence speed: where logarithmic inputs show an almost complete
convergence after a single epoch, linear inputs converge much
slower.\footnote{For lower learning rates, the fast convergence of
  logarithmic inputs is still observed, but linear inputs converge
  even slower.}
We also see that using ratios of \nsub as inputs yields a clearly
worse AUC.
Secondly, we can study convergence as a function of the size of the
network which can be either the network width at fixed number of
hidden layers, Fig.~\ref{fig:auc-v-width}, or the network depth at fixed
number of neurons per hidden layer, Fig.~\ref{fig:auc-v-depth}.
These plots also show that logarithmic inputs converge (marginally) faster
than linear inputs.
Interestingly, after a full training (with 7 hidden layers and 128
neurons per hidden layer), we recover a situation where logarithmic
inputs give a slightly better performance compared to linear inputs,
as our analytic studies suggested.

\begin{figure}
  \centering
  \begin{subfigure}[t]{0.48\textwidth}
    \includegraphics[width=\textwidth]{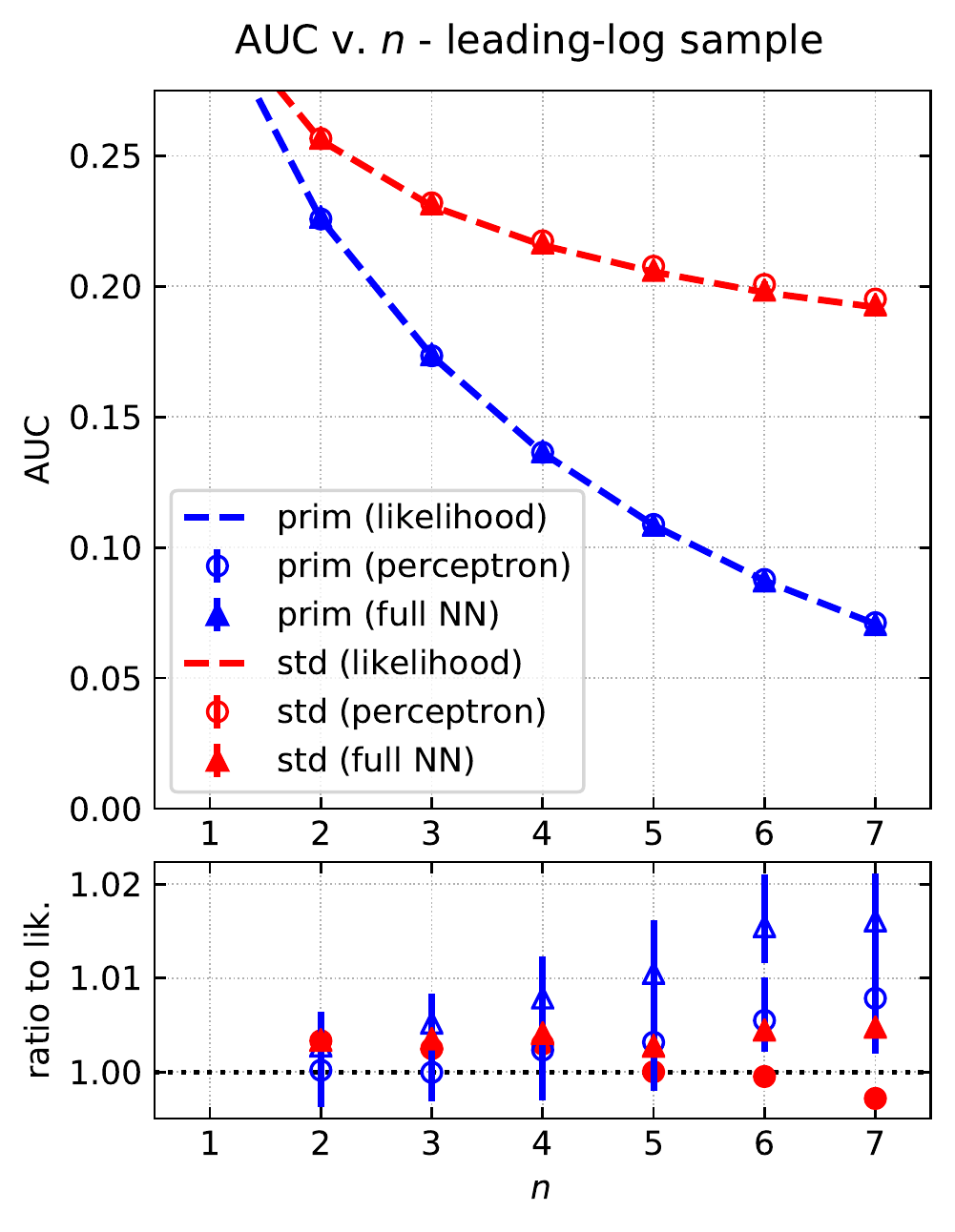}
    \caption{AUC as a function of the number of inputs $n$ for a
      leading-logarithmic distribution.}\label{fig:auc-v-n-ll}
  \end{subfigure}
  \hfill
  \begin{subfigure}[t]{0.48\textwidth}
    \includegraphics[width=\textwidth]{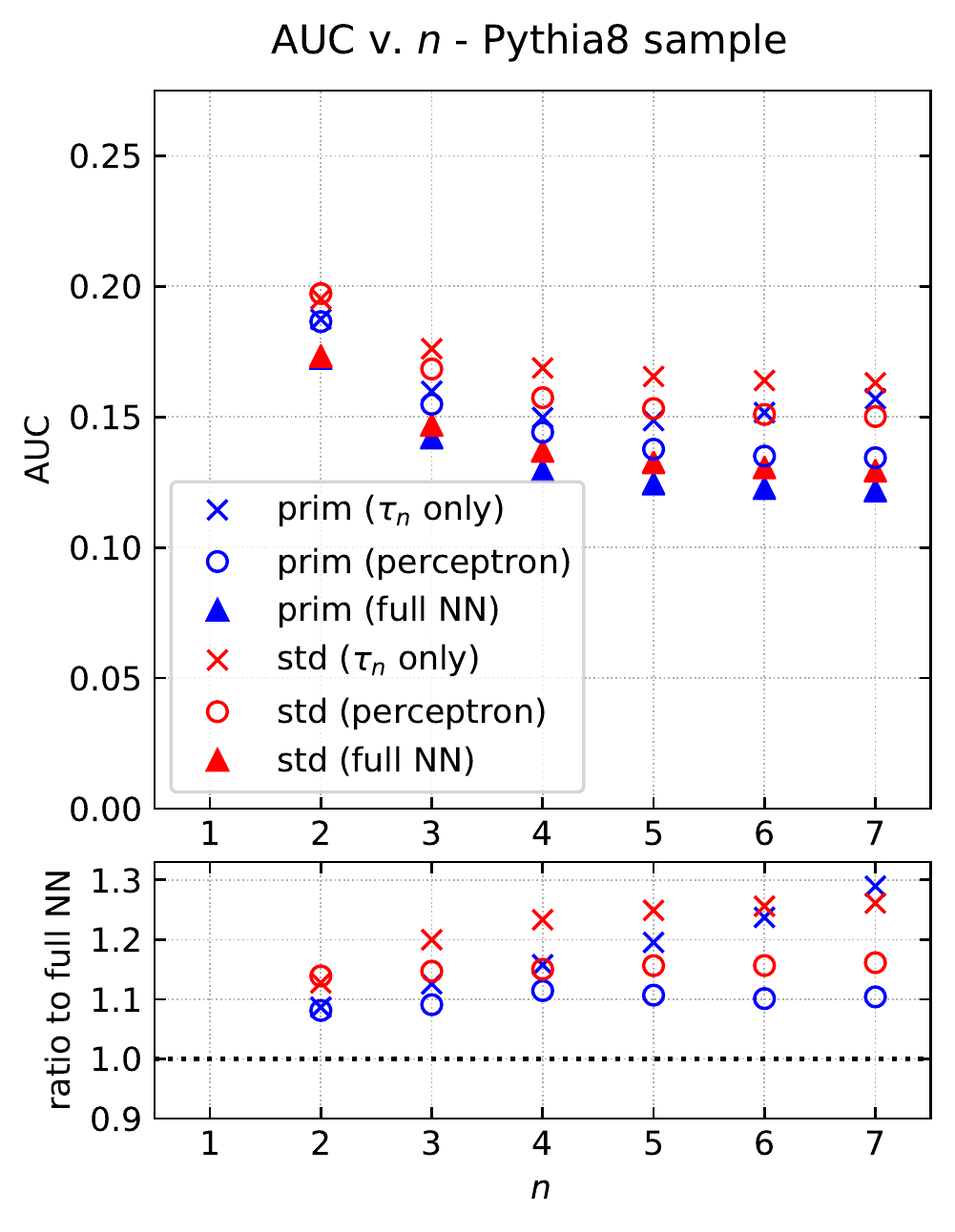}
    \caption{AUC as a function of the number of inputs $n$ for a
      Pythia8 distribution.}\label{fig:auc-v-n-pythia}    
  \end{subfigure}
  \caption{AUC as a function of the number of inputs $n$. We show
    results for both the standard (red) and primary (blue) definitions
    of \nsub with training done either for a single perceptron (open
    circles) or for a full NN (filled triangles). In all cases,
    log-square inputs have been used. When available, results for the
    likelihood have been included (dashed lines) and used to normalise
    the results in the lower panel. When not available, the
    normalisation is to the full NN results.
    For the Pythia8 sample we also show the results of a cut on
    $\tau_n$ (or $\taup_n$) only (crosses). 
  }\label{fig:auc-v-n}
\end{figure}

As a final test, we study in Fig.~\ref{fig:auc-v-n} the AUC as a function of the number $n$ of input \nsub
variables.
Results are shown for inputs distributed according to the
leading-logarithmic QCD distributions (as for our analytic calculations in
section~\ref{sec:analytics} and for our tests in
section~\ref{sec:numerics}) in Fig.~\ref{fig:auc-v-n-ll} as well as
for a Pythia8 sample in Fig.~\ref{fig:auc-v-n-pythia}.
We show the AUC for both the standard definition of \nsub (red) 
and for our primary definition (blue), and in both cases we show the results
of training either a single perceptron (open symbols) or a full NN
(filled symbols).
In the case of the Pythia8 sample, we also show the AUC
  corresponding to a simple cut on either $\taup_n$ or $\tau_n$.
For the case of the leading-logarithmic distributions, we also provide
the expectations from the actual likelihood ratio (dashed lines). The
latter is obtained from Eq.~(\ref{auc-prim-res}) for the case of
primary \nsub and by numerically integrating the equivalent
of Eq.~(\ref{auc-eval}) for the standard definition.
The lower panel shows the relative performance, normalised to the likelihood, if it is available, or to the full NN otherwise.
These results confirm once more our earlier findings.
First, primary \nsub performs better than the standard
definition, although the difference, clearly visible in the
leading-log distributions, is only marginal with the Pythia8 sample.
Then, in the case of a leading-log distribution, the optimal
performance of the primary \nsub is already captured by a
single perceptron, while for the case of standard \nsub
only the full NN is able to reach a performance on par with the
likelihood expectation.
For the Pythia8 sample, the training of the full NN leads a reduction
of the AUC compared to a single perceptron for both \nsub
definitions.
The improvement is however smaller for the primary definition ($\sim
10\%$) than for the standard one ($\sim 15\%$).
This is likely a reminiscence of the behaviour observed at leading
logarithmic accuracy.
It is interesting to observe that the performance of a simple cut
  on $\taup_n$, which is optimal at LL, is comparable to the
  perceptron at small $n$, while it degrades as $n$ increases. This is
  most likely due to the fact that at large $n$ we become more
  sensitive to non-perturbative corrections and consequently our LL
  approximation is no longer sufficient. Conversely, a cut on $\tau_n$, while giving
  a performance comparable to that of the perceptron for $n=2$, degrades
  more rapidly at larger $n$.

\section{Conclusions and outlook}\label{sec:conclusions}
In this paper we have investigated the behaviour of a simple neural network made of just one neuron, in the context of a binary classification problem. The novelty of our study consists in exploiting the underlying theoretical understanding of the problem, in order to obtain firm analytical predictions about the perceptron performance. 

In particular, we have addressed the question of distinguishing
quark-initiated jets from gluon-initiated ones, exploiting a new
variant of the \nsub (IRC-safe) family of observables that
we dubbed primary \nsub. As the name suggests, these observables are only sensitive, to LL  accuracy, to emissions of soft gluons off the original hard parton, while standard \nsub also depends on secondary gluon splittings, already at LL.  
Thanks to the simple all-order behaviour of observables
$\{\taup_1\dots \taup_n\}$ we have been able to determine that the
optimal discriminant at LL, i.e.\ one which is monotonically related
to the likelihood ratio, is just a cut on $\taup_n$. We have also been
able to obtain analytic expressions for  the area under the ROC curve
and the area under the ROC curve which are standard figures of merit for a classifier's performance. 
Furthermore, we have found that, besides having nicer theoretical properties, $\taup_N$ typically outperforms $\tau_N$ as a quark/gluon tagger, in the range of quark efficiencies that is typically employed in phenomenological analyses. 

The central part of this study has been the analytic study of the LL behaviour of a perceptron that takes primary \nsub variables as inputs. We have considered a perceptron with a sigmoid as activation function and the cross-entropy as loss function. 
The main question we have raised is whether the values of the
perceptron weights at the minimum of the cost function correspond to
the aforementioned optimal LL classifier, i.e.\ a cut on the last
primary \nsub considered. We have been able to show analytically that
this depends on the actual functional form of the inputs fed to the
network. The perceptron is able to find the correct minimum if
logarithms (or square logarithms) of the \nsub are passed, but fails
to do so with linear inputs. This reflects the fact that the LL
distributions of the inputs are indeed logarithmic. These analytic
results are fully confirmed by a numerical implementation of the
perceptron, which has been trained on pseudo-data generated according
to a LL distribution. Furthermore, we have also found that, in the
case of linear inputs, the learning rate of the perceptron is
substantially slower than for logarithmic inputs.  
As a by-product, we were also able to find, in a few cases, closed analytic expressions for the network cost function. 

Finally, we have considered a more realistic framework for our
analysis. Firstly, we have trained the perceptron pseudo-data
generated with a general-purpose Monte Carlo parton shower, and we
have obtained qualitative agreement with the analytic study. Secondly,
we have observed that when considering a full neutral network, we
obtain a comparable performance with all the variants of the inputs,
with the possible exception of the \nsub ratios, which requires more
fine-tuning to be brought to the same performance.

Our work highlights in a quantitative way the positive role that
first-principle understanding of the underlying physical phenomena has in classification problems that employs ML techniques.
Even if a similar degree of performance is ultimately achieved with a full NN
regardless of the type of inputs, a knowledge of the underlying physics
is helpful to build a simpler network less dependent on a careful optimisation of the hyperparameters.
Furthermore, we stress that the use of theoretically well-behaved observables has allowed us to perform our analysis at a well-defined, and in principle improvable, accuracy. We believe that this is an important step towards the consistent use of ML in experimental   measurements that are ultimately aimed at a detailed comparison with theoretical predictions. 

Our current analysis suffers from the obvious limitations of being
conducted at LL and for just a very simple network. We believe that
going beyond LL accuracy for primary \nsub is not only interesting,
but desirable.
The first reason would be to see if the simple resummation structure
we see at LL accuracy survives at higher orders. 
Then, this very first analysis places primary \nsub as a promising
substructure observable.
One is therefore tempted to also investigate its performance in the
context of boosted vector boson or top tagging, although one might expect that these classification problems
benefit from constraints beyond primary emissions.
Furthermore, in the framework of our analytic study of ML techniques,
it would be interesting to see if at a higher accuracy one is still
able to find a simple combination of architecture, cost function and
network inputs that guarantees, analytically, optimal performance.  
More generally, we look forward to future work where perturbative QCD
helps designing better ML techniques with a comfortable degree of
analytic control.

\begin{acknowledgments}
We thank the organisers of the BOOST 2019 workshop at MIT for a vibrant scientific environment that sparked the discussion about this project.
We are all thankful to Matteo Cacciari for multiple interactions and
interesting discussions.
We thank Simone Caletti, Andrea Coccaro, Andrew Larkoski, Ben Nachman, and Jesse Thaler for helpful comments on the manuscript.
SM and GiS wish to thank IPhT Saclay for hospitality during the course of this work. 
The work of SM is supported by Universit\`a di Genova under the curiosity-driven grant "Using jets to challenge the Standard Model of particle physics" and by the Italian Ministry of Research (MIUR) under grant PRIN 20172LNEEZ.
GK acknowledges support by the Deutsche Forschungsgemeinschaft (DFG, German Re\-search Foundation) 
under Germany's Excellence Strategy -- EXC 2121 "Quantum Universe" -- 390833306. 
GrS's work has been supported in part by the French Agence Nationale
de la Recherche, under grant ANR-15-CE31-0016.
\end{acknowledgments}

\appendix

\section{Details of the analytic calculations}\label{app:details}

In this appendix we report details of the analytic calculation of the area under the ROC curve (AUC).
We start from the generic definition of the AUC and we exploit the definition of the ROC curve given in Eq.~(\ref{roc-def}):
\begin{equation}
  \text{AUC} = \int_0^1 dx\, \text{ROC}(x)
  = \int_0^1 dx\, \Sigma_\BAC\left (\Sigma_\SIG^{-1}\left(x \right)\right),
\end{equation}
where, for a given observable $V$, the cumulative distribution for signal or
background as a function of a cut $\vcut$ reads as in Eq.~(\ref{eq:sigma-def}),
which we rewrite as:
\begin{equation}
  \Sigma_i(\vcut) = \int d\mathbf{t} \, p_i(\mathbf{t}) \,\Theta(V(\mathbf{t}) < \vcut),
  \quad i = \SIG, \BAC.
\end{equation}
We now perform a change of variable from the efficiency $x$ to the cut on the observable $\vcut$, which results in the following Jacobian factor:
\begin{equation}
  \frac{dx}{d\vcut} = \frac{dx}{d\Sigma_\SIG^{-1}\left(x \right)}
  = \frac{d\Sigma_\SIG\left(\vcut\right)}{d\vcut}.
\end{equation}
Thus, we have
\begin{align}
  \text{AUC} &= \int_0^1 d\vcut\, \Sigma_\BAC\left(\vcut\right) \frac{d\Sigma_\SIG\left(\vcut\right)}{d\vcut} \\
  &= \int_0^1 d\vcut\, \int d\mathbf{t}_\BAC \, p_\BAC(\mathbf{t}_\BAC) \,\Theta(V(\mathbf{t}_\BAC) < \vcut)
  \int d\mathbf{t}_\SIG \, p_\SIG(\mathbf{t}_\SIG) \,\delta(V(\mathbf{t}_\SIG) - \vcut) \\  
  &= \int d\mathbf{t}_\BAC \int d\mathbf{t}_\SIG \, p_\BAC(\mathbf{t}_\BAC) p_\SIG(\mathbf{t}_\SIG)
  \,\Theta(V(\mathbf{t}_\BAC) < v(\mathbf{t}_\SIG)),
\end{align}
which is the expression presented in Eq.~(\ref{auc-eval}).

We now specialise to our case and we consider the probability distributions $p_q$ and $p_g$ for primary \nsub at LL accuracy, which are given in Eq.~(\ref{eq:primdiff-end}).
We first integrate over quark variables and exploiting the result found in Eq.~(\ref{cumulative-res}), we obtain
\begin{equation}
  \int_0^1 d\taup_{1q}\cdots \int_0^{\taup_{n-1,q}} d\taup_{nq}
  \,p_q(\taup_{1q}, \taup_{2q}, \cdots, \taup_{nq})\,\Theta(\taup_{nq} > \taup_{ng})
  = 1 - \frac{\Gamma(n, \CF \Ralt(\taup_{ng}))}{\Gamma(n)}\,.
\end{equation}
The integral over $\taup_{ng}$ can be performed using the following integration-by-parts identity:
\begin{align}
  & \int_0^{\taup} d\taup' \,\CA \frac{\Ralt'(\taup')}{\taup'} e^{- \CA \Ralt(\taup')}\,\Gamma(n, \CF \Ralt(\taup')) \nonumber \\
  & \quad =  e^{- \CA \Ralt(\taup)} \Gamma(n, \CF \Ralt(\taup)) - \frac{\CF^n}{(\CF+\CA)^n} \Gamma(n, (\CF + \CA) \Ralt(\taup)) \,.
\end{align}
We see that iteratively the following kind of integrals appear: 
\begin{align}
  & \int_0^{\taup} d\taup' \,\CA \frac{\Ralt'(\taup')}{\taup}\,\Ralt(\taup')^m\,\Gamma(n, (\CF+\CA) \Ralt(\taup')) \nonumber \\
  & \quad = \frac{\CA}{1+m} \left( \frac{\Gamma(n+1+m, (\CF+\CA) \Ralt(\taup))}{(\CF+\CA)^{1+m}}
  -\Ralt(\taup)^{1+m} \Gamma(n, (\CF+\CA) \Ralt(\taup))
  \right)\,.
\end{align}
In the end we obtain:
\begin{align}
  & \int_0^1 d\taup_{1g} \cdots \int_0^{\taup_{n-1,g}} d\taup_{ng}
  \,p_g(\taup_{1g}, \taup_{2g}, \cdots, \taup_{ng})\,\Gamma(n, \CF \Ralt(\taup_{ng})) \nonumber \\
  & \quad = \Gamma(n) - \frac{\CF^n}{(\CF+\CA)^{2n-1}} \sum_{k=0}^{n-1} \frac{\Gamma(n+k)}{\Gamma(k+1)}
  \CA^k (\CF+\CA)^{n-1-k}.
\end{align}
The sum can be performed explicitly, leading to the result presented in Eq.~(\ref{auc-prim-res})\,.

\section{Analytic results for the cost function}\label{app:cost-function}

In section~\ref{sec:analytics} we have looked for a minimum of the cost function,
Eq.~(\ref{eq:costfun}), by computing from the very beginning the derivatives
with respect to the NN weights.
This procedure allowed us to determine analytically the position of the minimum for the case of log-square inputs. In the case of logarithmic and linear inputs, although we have not been able to perform all the integrals analytically, we have nonetheless reduced the problem of the determination of the position of the minimum of the cost function to a one-dimensional integration, which we have performed numerically. 

An alternative approach would be to determine an analytic expression for the cost function, as a function of the NN weights and bias, which we would have to, in turn, minimise. 
As we shall see in this appendix, this approach is not as successful as the one presented in the main text. The integrals that appear in the determination of the cost function Eq.~(\ref{eq:costfun}) are rather challenging and we have been able to solve them only in a couple of simple cases.
Namely, we limit ourselves to the case $n=2$ with log-square inputs, by using the sigmoid as activation function, and we report results for the cross-entropy loss, Eq.~(\ref{loss-xse}), and the quadratic loss, Eq.~(\ref{loss-chi2}).
Even if the NN setup is minimal, the derivation of explicit expressions
for the cost function may be instructive as they represent a first-principle determination of a NN behaviour and they could be valuable in the context of comparisons between experimental measurements that make use of machine-learning algorithms and theoretical predictions. 

\subsection{Cross-entropy loss}
We start by considering the cost function Eq.~(\ref{eq:costfun}) with the cross-entropy loss. Explicitly:
\begin{align}
  \CtildeXE(a_1, a_2, b) =& \frac{1}{2}\int_0^\infty d\logtsq_2 \int_0^{\logtsq_2} d\logtsq_1 \Bigg[
      \tCF^2\,e^{-\tCF \logtsq_2} \left( - \log\left(\frac{1}{1 + e^{a_1 \logtsq_1 + a_2\logtsq_2 + b}} \right)\right) \nonumber \\
      & \phantom{\int_0^\infty d\logtsq_2 \int_0^{\logtsq_2} d\logtsq_1 \Bigg[}
        + \tCA^2\,e^{-\tCA \logtsq_2} \left( - \log\left( \frac{1}{1 + e^{- a_1 \logtsq_1 - a_2\logtsq_2 - b}} \right) \right)
        \Bigg].
\end{align}
Note that the replacement $\tCF \leftrightarrow \tCA$ is equivalent to
$(a_1, a_2,b) \leftrightarrow (-a_1,-a_2,-b)$, due to the symmetries
of the functions involved. We first observe that the integral over $L_1$ gives rise to a dilogarithm.
In order to evaluate the integral over $L_2$ we make use of the following result:
\begin{align}
  & \intXE(\CR, c, d) = \CR^2 \int_0^\infty d x \, e^{-C_R x}\, \li{2}\left(-e^{c x+d} \right) \\
  &\quad = \CR \li{2}\left(- e^d\right) +
    \begin{cases} 
    -c \log\left(1+e^d\right) 
    -\frac{c^2}{\CR} \left( 1 - \hyp \left( 1, -\frac{\CR}{c}, 1-\frac{\CR}{c} ; -e^d
    \right)\right) &  (c<0)  \\ 
    0 & (c=0)  \\ 
    -c \log\left(1+e^d\right) 
    -\frac{c^2}{\CR} \, \hyp \left( 1, \frac{\CR}{c}, 1+\frac{\CR}{c} ; -e^{-d}\right)
    & (c>0)
  \end{cases} \nonumber
\end{align}
The analytic expression that we obtain for the cost function is 
\begin{align}\label{eq:ctildeXE}
  \CtildeXE(a_1, a_2, b) = \frac{1}{2a_1} \bigg[ &
    -(\intXE(\tCF, a_1+a_2, b) -  \intXE(\tCF, a_2, b)) \nonumber \\
    & +(\intXE(\tCA, - (a1+a_2), -b) - \intXE(\tCA, -a_2, -b)) \bigg]\,.
\end{align}
From section~\ref{sec:analytics}, we already now that the minimum is located at $a_1=0$. However, given the structure of the explicit result after integration,
 Eq.~(\ref{eq:ctildeXE}), it is highly nontrivial to recover the position of the minimum
analytically, due to the presence of $a_1$ both at denominator and in
the arguments of hypergeometric function.

\subsection{Quadratic loss}
\begin{figure}
  \centering
  \begin{subfigure}[t]{.49\textwidth}
    \includegraphics[width=\textwidth]{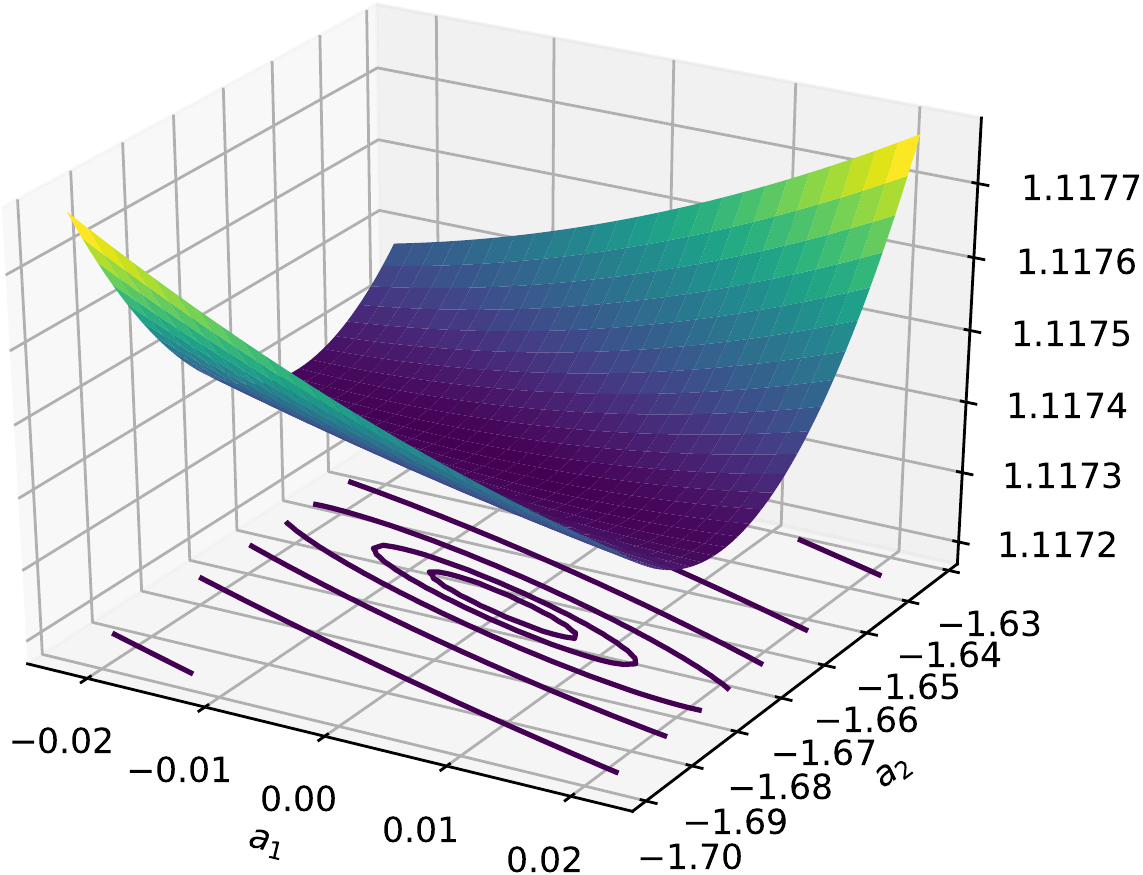}
    \caption{Cross-entropy loss}
  \end{subfigure}
  \hfill
  \begin{subfigure}[t]{.49\textwidth}
    \includegraphics[width=\textwidth]{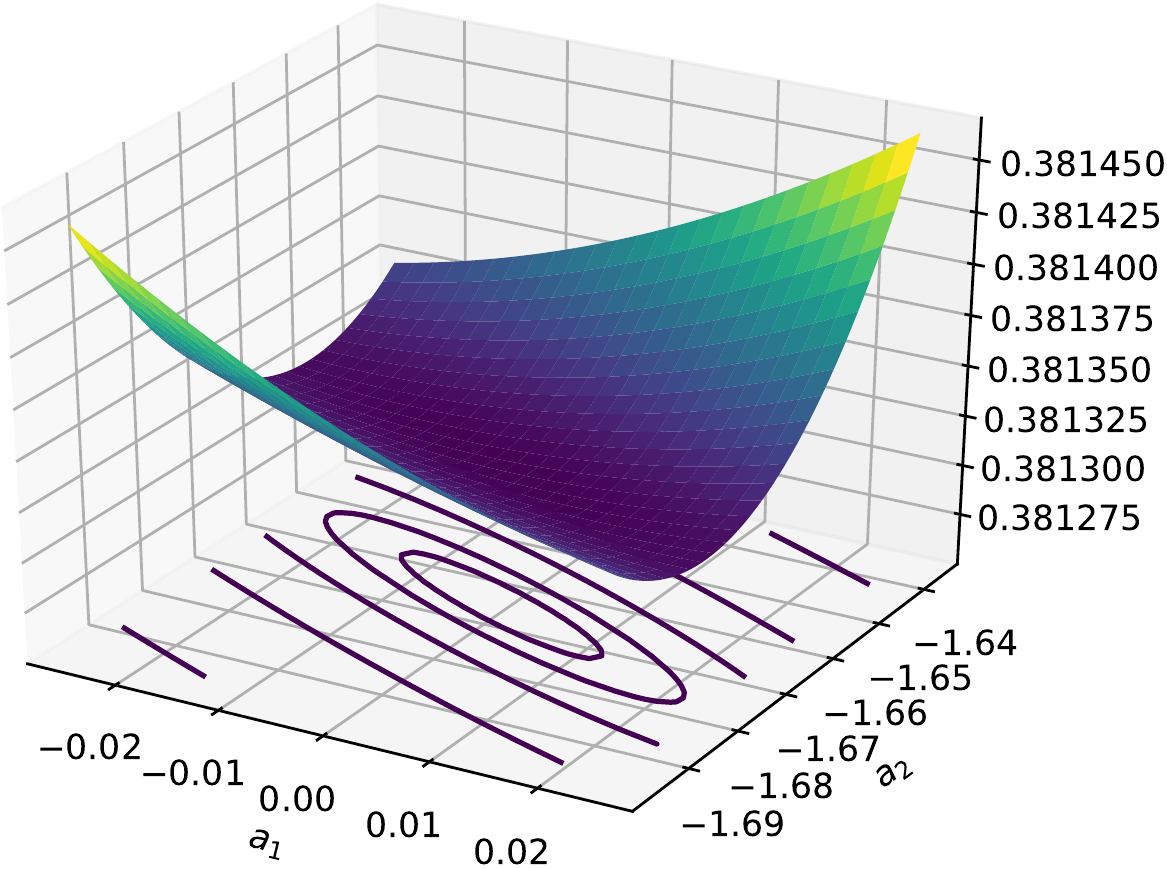}
    \caption{Quadratic loss}
  \end{subfigure}
  \caption{Cost function as a function of $a_1$ and $a_2$ around the minimum
    point.  $b$ has been accordingly fixed to the value in
    Eq.~(\ref{eq:minlogsq}).}\label{fig:costlogsq}
\end{figure}

An analogous calculation can be performed in the case of the quadratic loss Eq.~(\ref{loss-chi2}).
We have to calculate
\begin{equation}
  \CtildeCH(a_1, a_2, b) = \frac{1}{2}\int_0^\infty d\logtsq_2 \int_0^{\logtsq_2} d\logtsq_1 
  \Bigg[
    \frac{\tCF^2\,e^{-\tCF \logtsq_2}}{(1 + e^{- a_1 \logtsq_1 - a_2\logtsq_2 - b})^2} +
    \frac{\tCA^2\,e^{-\tCA \logtsq_2}}{(1 + e^{a_1 \logtsq_1 + a_2\logtsq_2 + b})^2} 
    \Bigg].
\end{equation}
As in the case of the cross-entropy loss, the integral over $L_1$ is straightforward.
We then make use of the following identity to perform the remaining integral over $L_2$:
\begin{align}
  & \intCH(\CR, c, d) = \CR^2 \int_0^\infty d x \, e^{-\CR x}\left[\log\left( 1+e^{cx+d}\right) +\frac{1}{1+e^{cx+d}}\right]\nonumber \\
  &\quad = \CR \log\left(1+e^d\right) + 
  \begin{cases}
    c + (\CR-c)\, \hyp \left( 1,-\frac{\CR}{c}, 1-\frac{\CR}{c} ; -e^{d}\right)
    & (c<0) \\
    \dfrac{\CR}{1+e^d}
    & (c=0) \\
    \CR + (c-\CR)\, \hyp \left( 1,\frac{\CR}{c} ,1+\frac{\CR}{c} ; -e^{-d}\right) 
    & (c>0) \\
  \end{cases}  
  \label{eq:quad-x2int}
\end{align}
We find
\begin{align}\label{eq:ctildeCH}
  \CtildeCH(a_1, a_2, b) = \frac{1}{2a_1} \bigg[ &
    \left( \intCH(\tCF, a_1+a_2, b) -  \intCH(\tCF, a_2, b) \right) \nonumber \\
    & - \left( \intCH(\tCA, - (a1+a_2), -b) - \intCH(\tCA, -a_2, -b) \right) \bigg].
\end{align}
As for Eq.~(\ref{eq:ctildeXE}), the position of the minimum is hard to find
analytically. To check whether we recover the result obtained in
Eq.~(\ref{eq:minlogsq}), we can numerically look for the minimum of
Eqs.~(\ref{eq:ctildeXE}) and (\ref{eq:ctildeCH}). For illustration purposes, we
fix $\tCA$ and $\tCF$ to $\CA$ and $\CF$ respectively. In
Fig.~\ref{fig:costlogsq} we plot each cost function around the found minimum
point. We see that the condition $a_1 = 0$ and the (negative) value of $a_2 =
\CF - \CA \simeq -1.67$ are indeed confirmed.

\section{Scaling of $a_n$ in the log inputs case.} \label{app:scaling}

In this appendix we would like to study the scaling of $a_n$ with respect to
$\tCF$ and $\tCA$. In order to simplify the notation, in this appendix we will
rename $x \equiv \logt_n$ and $a \equiv a_n$.

We first expand in series around $x=0$ the sigmoid functions which appear in
Eq.~(\ref{eq:Cderlog}):
\begin{align}
  \frac{1}{1 + e^{-a\,x - b}} &=
  \frac{e^b}{1+e^b} + \sum_{k=1}^{\infty} a^k\,f^{(k)}(e^b)\,x^k\,, \label{eq:ss1}\\
  -\frac{1}{1 + e^{a\,x + b}} &=
  -\frac{1}{1+e^b} + \sum_{k=1}^{\infty} a^k\,f^{(k)}(e^b)\,x^k\,, \label{eq:ss2}
\end{align}
where:
\begin{equation}
  f^{(k)}(e^b) = \frac{(-1)^{1+k}\,e^b}{\Gamma(1+k)(1+e^b)^{1+k}}
  \times \text{polynomial in $e^b$ with degree of $k-1$\,.}
\end{equation}
Note that the two series with index $k$ appearing on the rhs of
Eqs.~(\ref{eq:ss1})-(\ref{eq:ss2}) are the same.
By substituting Eqs.~(\ref{eq:ss1})-(\ref{eq:ss2}) under integration,
Eq.~(\ref{eq:Cderlog}) becomes:
\begin{align}
  \int_0^{\infty} dx\,x^m
  &\left[
    \tCF^n\,e^{-\tCF x^2}
    \left( \frac{e^b}{1+e^b} + \sum_{k=1}^{\infty} a^k\,f^{(k)}(e^b)\,x^k \right) \right. \nonumber \\
    &\quad \left. +\,\tCA^n\,e^{-\tCA x^2}
    \left( -\frac{1}{1+e^b} + \sum_{k=1}^{\infty} a^k\,f^{(k)}(e^b)\,x^k \right)
    \right] = 0
\end{align}
with $m=2n$ or $m=2n-1$.
Given the following integration identity:
\begin{equation}
  \int_0^\infty dx\,x^m\,e^{-c\,x^2} = \frac{1}{2\sqrt{c^{m+1}}} \Gamma\left(\frac{m+1}{2}\right), \quad c>0
\end{equation}
we obtain:
\begin{align}
  \frac{\Gamma\left(\frac{m+1}{2}\right)}{2(1+e^b)}
  &  \left( \frac{\tCF^n\,e^b}{\sqrt{\tCF^{m+1}}}
  - \frac{\tCA^n}{\sqrt{\tCA^{m+1}}} \right) \nonumber \\
  & + \sum_{k=1}^{\infty} a^k\,f^{(k)}(e^b)\,\frac{\Gamma\left(\frac{m+1+k}{2}\right)}{2}
  \left(\frac{\tCF^n}{\sqrt{\tCF^{m+1+k}}} + \frac{\tCA^n}{\sqrt{\tCA^{m+1+k}}}\right) = 0\,.
\end{align}
We now divide by $\sqrt{\tCF^{2n-m-1}}$ and collect $1/\sqrt{\tCF^k}$ in the second bracket:
\begin{align}
  \frac{\Gamma\left(\frac{m+1}{2}\right)}{2(1+e^b)}
  & \left( e^b  - \left(\frac{\tCA}{\tCF}\right)^{(2n-m-1)/2} \right) \nonumber \\
  & + \sum_{k=1}^{\infty} \left(
  \frac{a}{\sqrt{\tCF}}\right)^k\,f^{(k)}(e^b)\,\frac{\Gamma\left(\frac{m+1+k}{2}\right)}{2}
  \left( 1
  + \left(\frac{\tCA}{\tCF}\right)^{(2n-m-1-k)/2} \right) = 0\,.
\end{align}
This equation suggests that:
\begin{equation}
  \frac{a_n}{\sqrt{\tCF}} = F\left(\frac{\tCA}{\tCF}\right).
\end{equation}
which is the relation quoted in the main text.

\bibliography{biblio}

\end{document}